# Doubly Robust Smoothing of Dynamical Processes via Outlier Sparsity Constraints

Shahrokh Farahmand, *Student Member, IEEE*, Georgios B. Giannakis, *Fellow, IEEE* (Corresponding author), Daniele Angelosante, *Member, IEEE*

**Abstract**

Coping with outliers contaminating dynamical processes is of major importance in various applications because mismatches from nominal models are not uncommon in practice. In this context, the present paper develops novel fixed-lag and fixed-interval smoothing algorithms that are robust to outliers simultaneously present in the measurements *and* in the state dynamics. Outliers are handled through auxiliary unknown variables that are jointly estimated along with the state based on the least-squares criterion that is regularized with the $\ell_1$-norm of the outliers in order to effect sparsity control. The resultant iterative estimators rely on coordinate descent and the alternating direction method of multipliers, are expressed in closed form per iteration, and are provably convergent. Additional attractive features of the novel doubly robust smoother include: i) ability to handle both types of outliers; ii) universality to unknown nominal noise and outlier distributions; iii) flexibility to encompass maximum a posteriori optimal estimators with reliable performance under nominal conditions; and iv) improved performance relative to competing alternatives at comparable complexity, as corroborated via simulated tests.

The authors are with the Dept. of Electrical and Computer Engr. at the Univ. of Minnesota, Minneapolis, MN 55455, USA; (e-mails: {shahrokh, georgios,angel129}@umn.edu). This work was supported by NSF grants CCF 1016605 and CON 1002180. Part of this paper was presented at the *Asilomar Conf. on Signals, Systems, and Computers*, Pacific Grove, CA, Nov. 2010.

## I. Introduction

Estimating the state of dynamical systems is of paramount importance in various applications including tracking and navigation. A major challenge in these applications is deviation from nominal conditions, which gives rise to outliers in the observations and state dynamics. Outliers in the state may come from abrupt changes in the target position due to, e.g., unexpected turbulence, and velocity variations due to target maneuvering. Outliers in the observations typically occur because of clutter, and glint noise [26]. In addition, both types of outliers can arise after linearizing the emergent nonlinearities, as in the extended Kalman filter (EKF) [4], [24]. The clairvoyant Kalman filter (KF) and smoother (KS) can not handle state and/or measurement outliers [3], [27], because both can be viewed as minimizers of a weighted least-squares (WLS) criterion, which is known to be sensitive to outliers [22].

Robustification of KF and KS dates back to the ’70s [27], but remains an active area of research until today [33], [35], continuously leveraging advances in convex optimization [3], [5]. Despite these advances, existing robust KF and KS approaches have several limitations. Some consider outliers only in the measurements [33], while others can handle either type of outliers alone but not both simultaneously [27]. Most approaches capitalize on robust e.g., M-estimators [35], which rely on Huber’s and other outlier-resilient criteria [20, App. A6.8]. They require knowledge of the nominal distribution, and are effective only when the nominal noise is independent across observations and state entries [23, Chap. 7]. In the presence of correlated Gaussian noise, pre-whitening yields independent noise entries, which is required for M-estimates to be applicable [35]. However, pre-whitening spreads the outliers to non-contaminated measurements. Approaches to doubly robust fixed-lag smoothing rely on heuristics to determine whether outliers are present in the state or the measurement equation [35].

A recent scheme for robust fixed-interval (but not fix-lag) smoothing is reported in [3], treating nonlinearities in the state and measurement equations separately from robustness issues. In the development, nonlinearities are linearized, and the measurement noise is assumed to follow the so-termed $\ell_1$-Laplacian (or a Huber) distribution parameterized by a matrix $R$. The choice of $R$ (and likewise that of Huber thresholds) critically affects smoothing performance, but systematic means of selecting these parameters was left open in [3]. Finally, a class of robust schemes popular in computer vision for linear regression settings comprises the so-termed random sample consensus (RANSAC)-based algorithms [14], [20].

If the outlier distributions are known and the model is linear and Gaussian (when conditioned on the outliers), efficient sequential Monte Carlo (SMC) smoothers based on Rao-Blackwellization [9] as well as deterministic algorithms based on pruning techniques, such as the interacting multiple model (IMM) method [11], will offer viable alternatives. Unfortunately, accurate description of the outlier distribution can be hard to obtain in practice. In addition, the complexity of SMC methods can be prohibitive for medium-to-large size problems due to the curse of dimensionality [12].

In the present work, outliers are handled through auxiliary unknown variables that are *jointly* estimated along with the state. The resultant estimators rely on constraining the degree of outlier scarcity through $\ell_1$-norm regularization, which is imposed on the auxiliary variables to enable *sparsity* control. The proposed robust smoothers: i) can handle both types of outliers simultaneously (hence referred to as doubly robust); ii) are universal, meaning they can operate even when the distributions of the nominal noise and outliers are unknown; iii) possess maximum a posteriori (MAP) optimality under specific assumptions on the outlier and nominal noise distributions; iv) perform well under nominal conditions (i.e., with no outliers present); and v) outperform RANSAC- and Huber-based robust smoothers.

Unlike ordinary KS, the novel robust estimators are nonlinear functions of the data, and rely on the alternating direction method of multipliers (AD-MoM) or coordinate descent iterations. Closed-form expressions render the bulk of complexity per iteration comparable to that of KS, which is linear in the observation time. Few iterations of the coordinate descent or AD-MoM-based algorithms are required in practice to obtain satisfactory results. Numerical tests demonstrate that the developed methods can reject state and measurement outliers, and outperform RANSAC and Huber-based techniques.

The rest of the paper is organized as follows. Section II contains preliminaries and the problem statement. Fixed-interval doubly robust smoothing (DRS) is introduced in Section III, where the link between robustness and sparsity is also established. Selection of the regularization parameters is the subject of Section IV. The coordinate descent based DRS algorithm is developed in Section V. Fixed-lag DRS is dealt with in Section VI. An alternative formulation for general linear state-space models is developed in Section VII. Simulations are presented in Section VIII, and conclusions in Section IX.

*Notation:* Column vectors (matrices) are denoted with lower- (upper-) case boldface letters; $(\cdot)^T$ stands for transposition; $\mathbf{0}_N$ is the $N \times 1$ column vector with all zeros; and $\mathbf{I}_N$ is the $N \times N$ identity matrix. Given a set $\mathcal{S} \subset \mathbb{R}^N$, the indicator function is defined as $\mathbb{1}_{\mathcal{S}}(\mathbf{x}) = 1$ if $\mathbf{x} \in \mathcal{S}$, and $\mathbb{1}_{\mathcal{S}}(\mathbf{x}) = 0$, otherwise.

## II. PROBLEM STATEMENT AND PRELIMINARIES

Consider the following *outlier-aware* state-space model

$$\mathbf{x}_n = \mathbf{F}_n\mathbf{x}_{n-1} + \mathbf{w}_n + \mathbf{o}_{x,n}, \quad n = 1, \ldots, N \tag{1a}$$

$$\mathbf{y}_n = \mathbf{H}_n\mathbf{x}_n + \mathbf{v}_n + \mathbf{o}_{y,n}, \quad n = 1, \ldots, N \tag{1b}$$

where $\mathbf{x}_n \in \mathbb{R}^{D_x}$ and $\mathbf{y}_n \in \mathbb{R}^{D_y}$ denote the state and measurement vectors at time $n$, respectively; $\mathbf{w}_n$ and $\mathbf{v}_n$ are mutually independent, zero-mean nominal noise vectors, each independent across time, and from the initial state $\mathbf{x}_0$, with respective covariance matrices $\{\mathbf{Q}_n, \mathbf{R}_n\}_{n=1}^N$; $\mathbf{x}_0$ has mean $\mathbf{m}_0$ and covariance $\mathbf{\Sigma}_0$; and $\{\mathbf{o}_{x,n}, \mathbf{o}_{y,n}\}_{n=1}^N$ represent the unknown state and measurement outlier vectors.

Given $\{\mathbf{F}_n, \mathbf{H}_n, \mathbf{Q}_n, \mathbf{R}_n, \mathbf{y}_n\}_{n=1}^N$, $\mathbf{m}_0$, and $\mathbf{\Sigma}_0$, the goal of *fixed-interval* DRS is to estimate $\{\mathbf{x}_n\}_{n=1}^N$ and $\{\mathbf{o}_{x,n}, \mathbf{o}_{y,n}\}_{n=1}^N$. Different from [3], [27], [33], [35], note that the outliers are explicitly introduced and treated as unknown variables to be estimated. This problem can be cast as one of linear regression, since $\mathbf{x}_n = \mathbf{F}_n\mathbf{x}_{n-1} + \mathbf{o}_{x,n} + \mathbf{w}_n$ can be viewed as an extra "zero measurement" $\mathbf{0} = -\mathbf{x}_n + \mathbf{F}_n\mathbf{x}_{n-1} + \mathbf{o}_{x,n} + \mathbf{w}_n$; and similarly for the initial condition as $-\mathbf{m}_0 = -\mathbf{x}_0 + \mathbf{w}_0$, where $\mathbf{w}_0$ is zero-mean with covariance $\mathbf{\Sigma}_0$. Thus, (1) can be expressed in a matrix-vector form as

$$\left[\begin{array}{ccccc} -\mathbf{I} & & & & \\ \mathbf{F}_1 & -\mathbf{I} & & & \\ & \ddots & \ddots & & \\ & & \mathbf{F}_N & -\mathbf{I} & \\ \hline \mathbf{0} & \mathbf{H}_1 & & & \\ \vdots & & \ddots & & \\ \mathbf{0} & & & \mathbf{H}_N & \end{array}\right] \begin{bmatrix} \mathbf{x}_0 \\ \mathbf{x}_1 \\ \mathbf{x}_2 \\ \vdots \\ \mathbf{x}_N \end{bmatrix} + \left[\begin{array}{c} \mathbf{0} \\ \mathbf{o}_{x,1} \\ \vdots \\ \mathbf{o}_{x,N} \\ \hline \mathbf{o}_{y,1} \\ \vdots \\ \mathbf{o}_{y,N} \end{array}\right] + \left[\begin{array}{c} \mathbf{w}_0 \\ \mathbf{w}_1 \\ \vdots \\ \mathbf{w}_N \\ \hline \mathbf{v}_1 \\ \vdots \\ \mathbf{v}_N \end{array}\right] = \left[\begin{array}{c} -\mathbf{m}_0 \\ \mathbf{0} \\ \vdots \\ \mathbf{0} \\ \hline \mathbf{y}_1 \\ \vdots \\ \mathbf{y}_N \end{array}\right] \tag{2}$$

or in a more compact form (with obvious definitions) as

$$\mathbf{A}\mathbf{x} + \mathbf{o} + \mathbf{w} = \mathbf{y} \tag{3}$$

where matrix $\mathbf{A}$ is tall, and vector $\mathbf{w}$ has block diagonal covariance matrix $\mathbf{Q}_w :=$ diag$(\mathbf{\Sigma}_0, \mathbf{Q}_1, \ldots, \mathbf{Q}_N,$ $\mathbf{R}_1, \ldots, \mathbf{R}_N)$. Since both $\mathbf{x}$ and $\mathbf{o}$ are unknown, the linear system in (2) is clearly under-determined.

When there are no outliers (cf. $\mathbf{o} = \mathbf{0}$) and $\mathbf{A}$ is full rank, the WLS estimate [cf. (3)] $\hat{\mathbf{x}} := \arg\ \min_{\mathbf{x}}(\mathbf{y} -$ $\mathbf{A}\mathbf{x})^T\mathbf{Q}_w^{-1}(\mathbf{y}-\mathbf{A}\mathbf{x})$ yields the KS. Substituting from (2), this estimate can also be written as [1, p. 189]

$$\hat{\mathbf{x}}^{\text{KS}} := \arg\min_{\mathbf{x}} \ \frac{1}{2}\sum_{n=1}^{N}\|\mathbf{y}_n-\mathbf{H}_n\mathbf{x}_n\|^2_{\mathbf{R}_n^{-1}}+\frac{1}{2}\|\mathbf{x}_0-\mathbf{m}_0\|^2_{\mathbf{\Sigma}_0^{-1}}+\frac{1}{2}\sum_{n=1}^{N}\|\mathbf{x}_n-\mathbf{F}_n\mathbf{x}_{n-1}\|^2_{\mathbf{Q}_n^{-1}} \tag{4}$$

where $\|\mathbf{x}\|^2_{\mathbf{M}} := \mathbf{x}^T\mathbf{M}\mathbf{x}$. The estimate $\hat{\mathbf{x}}^{\text{KS}}$ is also known as the Rauch-Tung-Striebel (RTS) smoother [30]. It is both minimum mean-square error (MMSE) and maximum a posteriori (MAP) optimal if the initial state and all nominal noise vectors are Gaussian; otherwise, it is linear (L)MMSE optimal. In fact, adding to the WLS cost in (4) a ridge regularization term $\lambda\|\mathbf{x}\|_2^2$ to constrain the $\ell_2$ norm of $\mathbf{x}$, the resultant ridge WLS, as well as the (L)MMSE and MAP, all yield a unique estimate (even for under-determined models), and can be rendered equivalent depending on the assumptions and corresponding optimality claims one is willing to make. The exposition henceforth is centered around the (regularized) WLS approach, because it is universal with respect to (wrt) the underlying probability density functions.

With $\mathbf{o}=\mathbf{0}$ (or known for that matter), the state can be clearly estimated by solving the equations $(\mathbf{A}^T\mathbf{Q}_w^{-1}\mathbf{A})\mathbf{x} = \mathbf{A}^T\mathbf{Q}_w^{-1}(\mathbf{y}-\mathbf{o})$, where the matrix $\mathbf{A}^T\mathbf{Q}_w^{-1}\mathbf{A}$ has a *block tridiagonal* structure [cf. (2) and (3)]. This allows obtaining the solution in batch form at complexity which is linear in $N$ [19, p. 174]. Alternatively, one can use the forward-backward algorithm in e.g., [1, p. 189] or [30] to solve (4) recursively. The forward direction is a KF followed by the backward run, which smooths the filtered estimates. The forward-backward algorithm also exhibits linear complexity in $N$. In a nutshell, both batch and recursive solvers of (2)-(4) exhibit low complexity (linear in $N$) when $\mathbf{o}$ is known.

If unknown outliers $\mathbf{o}$ are present in (3), and one chooses to ignore them and run a clairvoyant KS as if $\mathbf{o}$ were absent, the MSE performance will be poor because the (W)LS criterion is known to be severely affected by outliers [23]. This mandates dealing with the outliers in (3) explicitly – a challenge addressed in the next section by exploiting sparsity constraints on $\mathbf{o}$.

## III. Robustness by controlling Outlier Sparsity

The under-determinacy in (3) when $\mathbf{o}$ is unknown, raises non-uniqueness and thus state identifiability issues. Ridge WLS, (L)MMSE, and MAP estimators cannot recover the exact $\mathbf{x}$, a fact confirmed by the nominal-noise-free setup [cf. $\mathbf{w}=\mathbf{0}$ in (3)], where one faces an under-determined system of linear equations generally admitting infinite solutions. Key to addressing this issue is the degree of *sparsity* (number of nonzero entries) of the vector $\mathbf{o}$ – an attribute offering the potential for solving uniquely under-determined systems of linear equations, as established recently in the context of compressive sampling [10]. This motivates recovery of a controllably sparse estimate of $\mathbf{o}$ by effecting sparsity through an $\ell_0$-(pseudo)norm regularization term. Specifically, the proposed robust smoother aims at

$$[\hat{\mathbf{x}}, \hat{\mathbf{o}}] := \arg \min_{\mathbf{x},\mathbf{o}} \frac{1}{2}\|\mathbf{y} - \mathbf{A}\mathbf{x} - \mathbf{o}\|^2_{\mathbf{Q}_w^{-1}} + \lambda\|\mathbf{o}\|_0 \tag{5}$$

where the scalar $\lambda$ is used to control the degree of sparsity in $\mathbf{o}$. The level of outlier sparsity can be selected by tuning $\lambda$, and the outliers can then be estimated jointly with the state via (5). Unfortunately, the $\ell_0$-norm renders the problem non-convex and in fact NP-hard, which suggests a convex relaxation using the closest *convex* approximation to the $\ell_0$-norm, namely the $\ell_1$-norm [10], [37].

Using an $\ell_1$-norm regularization and defining $\mathbf{o}_x := [\mathbf{o}^T_{x,1}, \ldots, \mathbf{o}^T_{x,N}]^T$, $\mathbf{o}_y := [\mathbf{o}^T_{y,1}, \ldots, \mathbf{o}^T_{y,N}]^T$ , the novel DRS approach amounts to [cf. (1)-(5)]

$$\begin{aligned}[\widehat{\mathbf{x}}^{\mathrm{DRS}}, \widehat{\mathbf{o}}_x, \widehat{\mathbf{o}}_y] := \arg \min_{\mathbf{x},\mathbf{o}_x,\mathbf{o}_y} \Bigg\{ & \frac{1}{2}\sum_{n=1}^{N} \|\mathbf{y}_n - \mathbf{H}_n\mathbf{x}_n - \mathbf{o}_{y,n}\|^2_{\mathbf{R}_n^{-1}} + \frac{1}{2}\sum_{n=1}^{N} \|\mathbf{x}_n - \mathbf{F}_n\mathbf{x}_{n-1} - \mathbf{o}_{x,n}\|^2_{\mathbf{Q}_n^{-1}} \\ & + \frac{1}{2}\|\mathbf{x}_0 - \mathbf{m}_0\|^2_{\mathbf{\Sigma}_0^{-1}} + \sum_{n=1}^{N} [\lambda_x\|\mathbf{o}_{x,n}\|_1 + \lambda_y\|\mathbf{o}_{y,n}\|_1] \Bigg\}\end{aligned} \tag{6}$$

where $\lambda_x$ and $\lambda_y$ are introduced in (6) to allow individual control of sparsity levels in $\mathbf{o}_{x,n}$ and $\mathbf{o}_{y,n}$. Viewing the cost in (6) as a Lagrangian function, allows casting this *unconstrained* minimization problem as a constrained one. Indeed, sufficiency of the Lagrange multiplier theory implies that [6, Sec. 3.3.4]: using the solution $\widehat{\mathbf{o}}_x, \widehat{\mathbf{o}}_y$ of (6) for given multipliers $\lambda_x$, $\lambda_y \geq 0$ and letting $\tau_x := \|\widehat{\mathbf{o}}_x\|_1$, $\tau_y := \|\widehat{\mathbf{o}}_y\|_1$, the equivalent *constrained* minimization problem entails the WLS cost (quadratic terms in (6)) subject to the constraints $\|\mathbf{o}_x\|_1 \leq \tau_x$, and $\|\mathbf{o}_y\|_1 \leq \tau_y$. Note however, that $\lambda_x(\lambda_y)$ in (6), and likewise $\tau_x(\tau_y)$ in its constrained equivalent, are tuning parameters and not optimization variables.

The DRS state estimate in (6) can cope with outliers *jointly* present in the state and in the measurements. In addition, it is *universal* because it does not require knowing the distribution of the nominal noise or the outlier vectors. (The choice of $\lambda_x$ and $\lambda_y$ discussed in the next section will not follow from the distribution of a contaminating model but will be data driven.) Different from [2] and [39] which enforce sparsity in the state, DRS controls sparsity in the outliers to effect robustness. At this point, it is worth recalling that $\mathbf{o}$ in smoothing dynamical processes is indeed sparse, since it models abrupt changes (target maneuvers) in the state which cannot be too many in the analysis window, and glint noise giving rise to large-magnitude observations which occur rarely too. Having explained why it is meaningful to expect only few nonzero entries in $\mathbf{o}$, it is also useful to clarify that this is not necessary. (Simulated tests in Section VIII will allow for outlier contamination as high as $80\%$.) Although smoothing performance degrades as the number of nonzero entries in $\mathbf{o}$ increases, all the proposed approach needs is a handle on the percentage of outliers without requiring this percentage to be necessarily low.

The WLS cost can be also replaced by other functions (e.g., the $\ell_1$-norm of the error), and alternative regularization terms (e.g., the $\ell_2$-norm of the outliers) can be employed instead of, or, in addition to the $\ell_1$-norm [18]. Non-convex costs and regularizers are also possible, but they are not recommended as stand-alone solvers of (6) because they cannot guarantee convergence to the global optimum. In contrast, it will be seen in Sections V and VI that (6) and variants involving $\ell_1$ and $\ell_2$ norms can afford not only globally convergent but also computationally efficient solvers.

Having clarified that the ensuing developments will rely on (6), which is meaningful regardless of the $\{\mathbf{w}, \mathbf{v}, \mathbf{o}\}$ distributions, it is natural to ask the following question: Under what assumptions on these distributions can one claim MAP optimality of the resultant state and outlier estimators? The ensuing proposition (proved in Appendix A) asserts that this is possible if the nominal noise vectors are Gaussian and the additive outliers are known to be Laplacian distributed.

**Proposition 1.** *Suppose* $\mathbf{w}_n$ *and* $\mathbf{v}_n$ *are Gaussian distributed, mutually independent, and independent from* $\mathbf{o}_{x,n}$ *and* $\mathbf{o}_{y,n}$*, respectively. Furthermore, assume* $\mathbf{o}_{x,n}$ *has Laplacian distributed entries* $o_{x,n,d}$ *that are independent from past states, past state outliers, measurement outliers, and across different dimensions; that is,* $o_{x,n,d}$ *and* $o_{x,n,d'}$ *are independent for* $d \neq d'$*. Similarly,* $\mathbf{o}_{y,n}$ *has Laplacian distributed entries* $o_{y,n,d}$ *that are independent from past states, past measurement outliers, state outliers, and across different dimensions; then the estimators obtained as in* (6) *are MAP optimal.*

Albeit simple to prove, the usefulness of Proposition 1 is twofold: (a) it allows for a side-by-side comparison with the MAP optimality offered by the clairvoyant KS in (4); and (b) it positions the proposed approach in the context of related MAP-optimal schemes adopting $\ell_1$-error based smoothers; see e.g., [3] and references therein. Specifically, different from the multivariate Laplacian in Proposition 1 described by the two scalar $\lambda_x$ and $\lambda_y$ parameters, the $\ell_1$-Laplacian model in [3] entails a $D_y \times D_y$ matrix of parameters that are assumed known.

Next, robustness of the estimators (6) is established. Specifically, the ensuing proposition proved in Appendix B, shows that DRS subsumes Huber's M-estimator as a special case.

**Proposition 2.** *When* $\{\mathbf{Q}_n, \mathbf{R}_n\}_{n=1}^N$ *and* $\boldsymbol{\Sigma}_0$ *are all identity matrices, the DRS in* (6) *boils down to solving the following Huber M-estimator problem*

$$\widehat{\mathbf{x}} := \arg\min_{\mathbf{x}} \left\{ \sum_{d=1}^{D_y} \sum_{n=1}^{N} \rho_{\lambda_y}(y_{n,d} - \mathbf{h}_{n,d}^T \mathbf{x}_n) + \sum_{d'=1}^{D_x} \left[ (x_{0,d'} - m_{0,d'})^2 + \sum_{n=1}^{N} \rho_{\lambda_x}(x_{n,d'} - \mathbf{f}_{n,d'}^T \mathbf{x}_{n-1}) \right] \right\} \quad (7)$$

*where* $\mathbf{y}_n := [y_{n,1}, y_{n,2}, \ldots, y_{n,D_y}]^T$, $\mathbf{x}_n := [x_{n,1}, x_{n,2}, \ldots, x_{n,D_x}]^T$, $\mathbf{x} := [\mathbf{x}_1^T, \mathbf{x}_2^T, \ldots, \mathbf{x}_N^T]^T$, $\mathbf{H}_n := [\mathbf{h}_{n,1}, \ldots, \mathbf{h}_{n,D_y}]^T$, $\mathbf{F}_n := [\mathbf{f}_{n,1}, \ldots, \mathbf{f}_{n,D_x}]^T$, *and* $\rho_\lambda$ *denotes the Huber cost [23]*

$$\rho_\lambda(r) := \begin{cases} \frac{1}{2} r^2, & \textit{if } |r| \leq \lambda \\ \lambda |r| - \frac{\lambda^2}{2}, & \textit{otherwise.} \end{cases}$$

Proposition 2 generalizes to dynamical systems the link between (6) and (7) established in [17] for linear regression models. As a result, DRS also inherits the robustness attributes associated with the Huber M-estimators. Figure 1 depicts Huber's cost along with the quadratic one. For small residuals (i.e., $|r| \leq \lambda$), $\rho_\lambda(r)$ coincides with the quadratic one. But for $|r| > \lambda$, Huber's cost grows only linearly with $r$, which allows for down-weighting large errors. Therefore, outliers which are responsible for large errors will be weighted less in the overall objective function. Clearly, for large values of $\lambda$, Huber's cost coincides with the quadratic one. As a consequence, a large number of outliers in the observations and state is effected through small $\lambda_y$ and $\lambda_x$, respectively. Finally, it should be mentioned that the Huber function is not the only one enabling robustness. A gamut of related robust costs can be found in e.g., [20, Appendix A6.8] with different properties. The most convincing reason for exploiting sparsity constraints under the $\ell_1$-norm of the outlier vectors is to leverage recent advances on compressive sampling to develop the computationally efficient and globally convergent solvers presented in Section V.

While DRS inherits the robustness features of Huber's M-estimator, it enjoys several advantages over it, as detailed in the following two remarks.

**Remark 1.** As mentioned earlier, the universality of DRS pertains also to the regularization term. If outliers are present in *all* entries of $\mathbf{x}_n$ or $\mathbf{y}_n$, this form of *group sparsity* can be effected by replacing $\|\mathbf{o}_{y,n}\|_1$ and $\|\mathbf{o}_{x,n}\|_1$ in (6) with $\|\mathbf{o}_{y,n}\|_2$ and $\|\mathbf{o}_{x,n}\|_2$, respectively. Using the latter regularization, either $\mathbf{o}_{y,n} = \mathbf{0}_{D_y}$ ($\mathbf{o}_{x,n} = \mathbf{0}_{D_x}$), or *all* the entries of $\mathbf{o}_{y,n}$ ($\mathbf{o}_{x,n}$) are nonzero, signifying the presence of outliers in *all* measurement (state) variables at time $n$. The cost function resulting from $\ell_2$-norm regularization is still convex [41], and its minimization can be carried out using solvers similar to those of (6) to be presented in Sections V and VI. For this reason, only $\ell_1$-norm regularization will be considered henceforth.

**Remark 2.** In addition to universal robustness, the novel approach to DRS is also flexible in three counts. First, Huber estimators fix $\lambda_x$ or $\lambda_y$ *a fortiori* based on knowledge of the nominal distribution and the contamination model, e.g., for the $\epsilon$-contaminated class with Gaussian nominal, it follows that $\lambda_x = \lambda_y = 1.345$ [23]. In contrast, DRS does not assume any specific model for the outliers' distribution. From this viewpoint, $M$-estimators are subsumed by the present formulation as special cases corresponding to specific values of $\lambda_x$ and $\lambda_y$. In addition, DRS can accommodate colored noise [cf. (6)], which is formidable for the robust estimators of [27] and [35] because pre-whitening in (7) with $\mathbf{Q}_n^{-0.5}$ and $\mathbf{R}_n^{-0.5}$ spreads the outliers to non-contaminated measurements. Finally, DRS not only allows one to apply KS on outlier-free data but also reveals the outliers – a feature not available to Huber-based approaches, which only implicitly incorporate the outliers.

The next section presents systematic means of adjusting $\lambda_x$ and $\lambda_y$ to accommodate fully nominal settings (i.e., no outliers), fully contaminated scenarios, and all cases in between, even when the degree of contamination is unknown.

## IV. Selecting $\lambda_x$ and $\lambda_y$

Parameters $\lambda_x$ and $\lambda_y$ control the level of sparsity in the estimated outlier vectors, and their judicious selection is crucial for the successful operation of DRS. Too large a value for these parameters reverts DRS back to the KS, which is non-robust. On the other hand, very small values give rise to many spurious state and measurement outliers, thus degrading DRS performance. Standard cross-validation techniques [31], are not effective when outliers are present [25]. Toward choosing proper values of $\lambda_x$ and $\lambda_y$, the next proposition provides computable bounds so that if $\lambda_y \geq \bar{\lambda}_y$ and $\lambda_x \geq \bar{\lambda}_x$, then DRS coincides with KS. (See Appendix C for the proof.)

**Proposition 3.** *The DRS estimate in* (6) *coincides with KS estimate* $\widehat{\mathbf{x}}^{\text{KS}}$ *if*

$$\lambda_y \geq \bar{\lambda}_y \quad := \quad \max_{1\leq n\leq N} \left\|\mathbf{R}_n^{-1}(\mathbf{y}_n - \mathbf{H}_n\widehat{\mathbf{x}}_n^{\text{KS}})\right\|_\infty \tag{8a}$$

$$\lambda_x \geq \bar{\lambda}_x \quad := \quad \max_{1\leq n\leq N} \left\|\mathbf{Q}_n^{-1}(\widehat{\mathbf{x}}_n^{\text{KS}} - \mathbf{F}_n\widehat{\mathbf{x}}_{n-1}^{\text{KS}})\right\|_\infty . \tag{8b}$$

Having established the upper bounds in (8), desirable values for $\lambda_x$ and $\lambda_y$ will be points in the rectangle $[0, \bar{\lambda}_x] \times [0, \bar{\lambda}_y]$. Consider a two-dimensional grid on this rectangle and a properly chosen cost generated by each grid point. Depending on the information available to the designer, the "best" $\lambda_x$ and $\lambda_y$ will be those values minimizing either one of two costs presented in the ensuing subsections.

### *A. Known percentage of outliers*

Here the percentage of (non-)zero entries of the outlier vectors is assumed (at least approximately) known; denote them as $\pi_{o,x}$ and $\pi_{o,y}$. Consider the 2-D grid on $[0, \bar{\lambda}_x] \times [0, \bar{\lambda}_y]$, comprising $I_x$ points along the $\lambda_x$ axis and $I_y$ points along the $\lambda_y$ axis. Let $(i_x, i_y)$, with $1 \leq i_x \leq I_x$ and $1 \leq i_y \leq I_y$ be the grid point corresponding to values $\lambda_x(i_x)$ and $\lambda_y(i_y)$. For the given $(i_x, i_y)$, solve (6) with $\lambda_x = \lambda_x(i_x)$ and $\lambda_y = \lambda_y(i_y)$, to obtain $\widehat{\mathbf{x}}(i_x, i_y), \widehat{\mathbf{o}}_x(i_x, i_y)$ and $\widehat{\mathbf{o}}_y(i_x, i_y)$. With $\text{supp}(\mathbf{x})$ representing the non-zero entries of $\mathbf{x}$, and $|\mathbf{x}|$ the number of entries of $\mathbf{x}$, the "best" $\lambda_x(i_x^*)$ and $\lambda_y(i_y^*)$ are found as those with

$$[i_x^*, i_y^*] \quad := \quad \arg\min_{i_x, i_y} \left\{ \left| \pi_{x,o} - \frac{|\text{supp}(\widehat{\mathbf{o}}_x(i_x, i_y))|}{|\widehat{\mathbf{o}}_x(i_x, i_y)|} \right| + \left| \pi_{y,o} - \frac{|\text{supp}(\widehat{\mathbf{o}}_y(i_x, i_y))|}{|\widehat{\mathbf{o}}_y(i_x, i_y)|} \right| \right\} . \tag{9}$$

Finding $\lambda_x(i_x^*)$ and $\lambda_y(i_y^*)$ as in (9), appears to require solving (6) for all pairs $(i_x, i_y)$ of the two-dimensional grid. The associated computational cost can be viewed as the "price paid" for the universality attribute of DRS elaborated in Section III. Instead of two, $(\lambda_x, \lambda_y)$, recall that the number of parameters (and thus dimensionality of the search space had those been unknown) in [3] is $\mathcal{O}(D_y^2)$. Of course, this is not an issue in [3] where these parameters are assumed known.

Fortunately, the special structure of the optimization problem in (6) allows for solvers at complexity lower than running $I_x I_y$ robust smoothers, one per $(\lambda_x, \lambda_y)$ point on the grid. Indeed, (6) can be formulated as a quadratic program (QP), and its form can leverage recent advances in computing the so-termed least-absolute shrinkage and selection operator (Lasso), originally developed for static linear regressions; see e.g., [21]. As will be detailed in Section V, Lasso can be also exploited for the DRS dynamical model considered here. General-purpose QP solvers incur polynomial complexity up to $\mathcal{O}(D^{3.5})$ per iteration, where $D$ is the number of optimization variables involved [8]; here, $D = N(2D_x + D_y + 1)$. The reduction to $\mathcal{O}(D)$ per iteration afforded by Lasso-based solvers becomes possible by starting from $\lambda := (\bar{\lambda}_x, \bar{\lambda}_y)$ (sparsest initialization) and solving successively over decreasing $\lambda$-points on the grid, using coordinate descent iterations. Qualitatively speaking, about one nonzero entry of $\mathbf{o}$ emerges per $\lambda$-point on the grid, and its value is used to initialize the iteration for the next point on the grid (warm start) [15], [40]. Especially for large problem dimensions ($D \gg$), it has been demonstrated that such Lasso solutions for the entire so-termed *regularization path* (corresponding to all $\lambda$-points on the grid), can be more computationally efficient than solving Lasso even for a single fixed point $\lambda$ on the grid; see also [16].

### B. Known covariance of nominal noise vectors

The key observation here is that if the estimates $\widehat{\mathbf{o}}_{x,n}(i_x, i_y)$ and $\widehat{\mathbf{o}}_{y,n}(i_x, i_y)$ are accurate, then $\widehat{\mathbf{x}}_n(i_x, i_y) - \mathbf{F}_n\widehat{\mathbf{x}}_{n-1}(i_x, i_y) - \widehat{\mathbf{o}}_{x,n}(i_x, i_y)$ should have the same statistics as $\mathbf{w}_n$; and likewise, the statistics of $\mathbf{y}_n - \mathbf{H}_n\widehat{\mathbf{x}}_n(i_x, i_y) - \widehat{\mathbf{o}}_{y,n}(i_x, i_y)$ should coincide with those of $\mathbf{v}_n$. Focusing for instance on second-order statistics, if these estimated residuals are pre-whitened (by left-multiplication with $\mathbf{Q}_n^{-0.5}$ and $\mathbf{R}_n^{-0.5}$), they should have zero mean and unit variance. Thus, upon pre-whitening and averaging, their sample variance should approach 1. As a consequence, the "best" $\lambda_x(i_x^*)$ and $\lambda_y(i_y^*)$ are found as those with

$$
\begin{aligned}
[i_x^*, i_y^*] &:= \arg\min_{i_x, i_y} \left|1 - \hat{\sigma}_e^2(i_x, i_y)\right| \qquad (10)\\
\hat{\sigma}_e^2(i_x, i_y) &:= \frac{\|\widehat{\mathbf{w}}_0(i_x, i_y)\|^2_{\boldsymbol{\Sigma}_0^{-1}} + \sum_{n=1}^N \left[\|\widehat{\mathbf{v}}_n(i_x, i_y)\|^2_{\mathbf{R}_n^{-1}} + \|\widehat{\mathbf{w}}_n(i_x, i_y)\|^2_{\mathbf{Q}_n^{-1}}\right]}{ND_y + (N+1)D_x}
\end{aligned}
$$

where

$$
\begin{aligned}
\widehat{\mathbf{v}}_n(i_x, i_y) &:= \mathbf{y}_n - \mathbf{H}_n\widehat{\mathbf{x}}_n(i_x, i_y) - \widehat{\mathbf{o}}_{y,n}(i_x, i_y)\\
\widehat{\mathbf{w}}_n(i_x, i_y) &:= \widehat{\mathbf{x}}_n(i_x, i_y) - \mathbf{F}_n\widehat{\mathbf{x}}_{n-1}(i_x, i_y) - \widehat{\mathbf{o}}_{x,n}(i_x, i_y)\\
\widehat{\mathbf{w}}_0(i_x, i_y) &:= \widehat{\mathbf{x}}_0(i_x, i_y) - \mathbf{m}_0.
\end{aligned}
$$

The number of grid points $I_x$ and $I_y$ should be chosen large enough to ensure that a point in the vicinity of the global minimum of (10) is obtained. The grid need not be uniform. Indeed, simulations confirm that the search is more efficient if grid points are chosen on the log scale; see also [16]. This parameter tuning method, will be henceforth referred to as absolute variance deviation (AVD). Since DRS in (6) requires knowledge of nominal noise covariances, the AVD scheme needs no additional assumption; and similar to the method of the previous subsection, it can capitalize on Lasso coordinate descent based schemes to lower the computational complexity of solving (6) per grid point, as detailed next.

## V. DRS VIA COORDINATE DESCENT

While general purpose QP solvers can be utilized to solve (6) with polynomial complexity in $N$, their complexity can still be too high when $N$ is large. A reduced-complexity alternative is developed in this section to solve (6) using block coordinate descent iterations. Letting $C(\mathbf{x}, \mathbf{o}_x, \mathbf{o}_y)$ denote the cost in (6) and $j$ indexing coordinate descent iterations, the following sub-problems are solved per iteration $j$ and coordinate dimension $d$

$$\mathbf{x}^{(j)} = \arg\min_{\mathbf{x}} C(\mathbf{x}, \mathbf{o}_x^{(j-1)}, \mathbf{o}_y^{(j-1)}) \tag{11a}$$

$$o_{x,n,d}^{(j)} = \arg\min_{o_{x,n,d}} C(\mathbf{x}^{(j)}, \mathbf{o}_{x,1}^{(j)}, \ldots, \mathbf{o}_{x,n-1}^{(j)}, \mathbf{o}_{x,n,1:d-1}^{(j)}, o_{x,n,d}, \mathbf{o}_{x,n,d+1:D_x}^{(j-1)}, \mathbf{o}_{x,n+1}^{(j-1)}, \ldots, \mathbf{o}_{x,N}^{(j-1)}, \mathbf{o}_y^{(j-1)}) \tag{11b}$$

$$o_{y,n,d}^{(j)} = \arg\min_{o_{y,n,d}} C(\mathbf{x}^{(j)}, \mathbf{o}_x^{(j)}, \mathbf{o}_{y,1}^{(j)}, \ldots, \mathbf{o}_{y,n-1}^{(j)}, \mathbf{o}_{y,n,1:d-1}^{(j)}, o_{y,n,d}, \mathbf{o}_{y,n,d+1:D_y}^{(j-1)}, \mathbf{o}_{y,n+1}^{(j-1)}, \ldots, \mathbf{o}_{y,N}^{(j-1)}) \tag{11c}$$

where (11b) is solved for $n = 1, \ldots, N$ and $d = 1, \ldots, D_x$, while (11c) is solved for $n = 1, \ldots, N$ and $d = 1, \ldots, D_y$. The initial conditions are $\mathbf{o}_x^{(0)} = \mathbf{0}_{ND_x}$ and $\mathbf{o}_y^{(0)} = \mathbf{0}_{ND_y}$.

The optimization in (11a) can be explicitly written as

$$\mathbf{x}^{(j)} := \arg\min_{\mathbf{x}} \left\{ \frac{1}{2} \sum_{n=1}^{N} \left\| \mathbf{y}_n - \mathbf{H}_n \mathbf{x}_n - \mathbf{o}_{y,n}^{(j-1)} \right\|_{\mathbf{R}_n^{-1}}^2 \right. \\ \left. + \frac{1}{2} \sum_{n=1}^{N} \left\| \mathbf{x}_n - \mathbf{F}_n \mathbf{x}_{n-1} - \mathbf{o}_{x,n}^{(j-1)} \right\|_{\mathbf{Q}_n^{-1}}^2 + \frac{1}{2} \|\mathbf{x}_0 - \mathbf{m}_0\|_{\boldsymbol{\Sigma}_0^{-1}}^2 \right\}. \tag{12}$$

Solving (12) is equivalent to finding the KS estimate for a system with outlier-compensated measurements $\mathbf{y}_n - \mathbf{o}_{y,n}^{(j-1)}$, and outlier-compensated state $\mathbf{x}_n - \mathbf{o}_{x,n}^{(j-1)}$. Therefore, either the batch or the forward-backward recursive algorithms reviewed in Section II can be adopted to solve (12) with linear complexity in $N$.

Focusing on (11b), one should solve

$$o_{x,n,d}^{(j)} := \arg\min_{o_{x,n,d}} \frac{1}{2} \left\| \mathbf{x}_n^{(j)} - \mathbf{F}_n \mathbf{x}_{n-1}^{(j)} - \begin{bmatrix} \mathbf{o}_{x,n,1:d-1}^{(j)} \\ o_{x,n,d} \\ \mathbf{o}_{x,n,d+1:D_x}^{(j-1)} \end{bmatrix} \right\|_{\mathbf{Q}_n^{-1}}^2 + \lambda_x |o_{x,n,d}| \tag{13}$$

for every $d = 1, \ldots, D_x$ and $n = 1, \ldots, N$. The scalar problem (13) is solved using the Lasso, which can afford a closed-form solution [21]. Indeed, (13) can be equivalently expressed as

$$o_{x,n,d}^{(j)} := \arg\min_{o_{x,n,d}} \frac{1}{2} \left( o_{x,n,d} - \gamma_{x,n,d}^{(j)} \right)^2 + \lambda_{x,n,d} |o_{x,n,d}| \tag{14}$$

where

$$\gamma_{x,n,d}^{(j)} := \frac{1}{q_{n,d,d}} [\alpha_{x,n,d}^{(j)} - \sum_{k=1}^{d-1} q_{n,k,d} o_{x,n,k}^{(j)} - \sum_{k=d+1}^{D_x} q_{n,k,d} o_{x,n,k}^{(j-1)}]$$

$$\boldsymbol{\alpha}_{x,n}^{(j)} := \mathbf{Q}_n^{-1} \left( \mathbf{x}_n^{(j)} - \mathbf{F}_n \mathbf{x}_{n-1}^{(j)} \right), \qquad \lambda_{x,n,d} := \lambda_x / q_{n,d,d}$$

and $\mathbf{Q}_n^{-1}$ has entries $[\mathbf{Q}_n^{-1}]_{k,k'} := q_{n,k,k'}$. The solution to (14) is given by

$$o_{x,n,d}^{(j)} = \left[\left|\gamma_{x,n,d}^{(j)}\right| - \lambda_{x,n,d}\right]^{+} \text{ sign } \left(\gamma_{x,n,d}^{(j)}\right)$$

where $[x]^{+} := \max(x, 0)$ and sign$(\cdot)$ denotes the sign operator.

A similar closed-form solution becomes available for (11c), since

$$o_{y,n,d}^{(j)} := \arg\min_{o_{y,n,d}} \frac{1}{2} \left\| \mathbf{y}_n - \mathbf{H}_n \mathbf{x}_n^{(j)} - \begin{bmatrix} \mathbf{o}_{y,n,1:d-1}^{(j)} \\ o_{y,n,d} \\ \mathbf{o}_{y,n,d+1:D_y}^{(j-1)} \end{bmatrix} \right\|_{\mathbf{R}_n^{-1}}^{2} + \lambda_y |o_{y,n,d}|. \tag{15}$$

for every $d = 1, \ldots, D_y$ and $n = 1, \ldots, N$.

Problem (15) can be alternatively written as

$$o_{y,n,d}^{(j)} := \arg\min_{o_{y,n,d}} \frac{1}{2} \left(o_{y,n,d} - \gamma_{y,n,d}^{(j)}\right)^2 + \lambda_{y,n,d} |o_{y,n,d}| \tag{16}$$

where

$$\gamma_{y,n,d}^{(j)} := \frac{1}{r_{n,d,d}} [\alpha_{y,n,d}^{(j)} - \sum_{k=1}^{d-1} r_{n,k,d} o_{y,n,k}^{(j)} - \sum_{k=d+1}^{D_y} r_{n,k,d} o_{y,n,k}^{(j-1)}]$$

$$\boldsymbol{\alpha}_{y,n}^{(j)} := \mathbf{R}_n^{-1} \left(\mathbf{y}_n - \mathbf{H}_n \mathbf{x}_n^{(j)}\right), \qquad \lambda_{y,n,d} := \lambda_y / r_{n,d,d}$$

and $\mathbf{R}_n^{-1}$ has entries $[\mathbf{R}_n^{-1}]_{k,k'} := r_{n,k,k'}$. The solution to (14) is given by

$$o_{y,n,d}^{(j)} = \left[\left|\gamma_{y,n,d}^{(j)}\right| - \lambda_{y,n,d}\right]^{+} \text{ sign } \left(\gamma_{y,n,d}^{(j)}\right).$$

Global convergence of the (12)-(15) iterates is guaranteed from the results in [38], as summarized next.

**Proposition 4.** *For any initial values* $\mathbf{x}^{(0)}, \mathbf{o}_x^{(0)}, \mathbf{o}_y^{(0)}$*, the iterates in* (12), (13) *and* (15) *are all convergent. Furthermore, every limit point of the sequences* $\mathbf{x}^{(j)}$*,* $\mathbf{o}_x^{(j)}$*,* $\mathbf{o}_y^{(j)}$ *solves* (6).

Note that (12) contains the bulk of computation per iteration $j$, and its complexity is equivalent to that of KS, which is linear in $N$. This should be contrasted with the general purpose convex solvers whose complexity is polynomial in $N$ (worst-case of order $\mathcal{O}(N^{3.5})$; see e.g., [8]). As mentioned earlier, the complexity reduction is due to the unique properties of Lasso-related problems, namely variable separability, closed-form thresholding per variable, and warm starts. Coordinate descent solvers capitalize on these properties, and have been documented to outperform competing alternatives, including off-the-shelf QP solvers [15], [16], [40].

**Remark 3.** This section's efficient solvers of the $\ell_1$-norm based convex cost in (6) will converge to estimates generally not coinciding with the global optimum of the ultimate $\ell_0$-norm based sparsity-promoting cost in (5). This motivates *concave* regularization terms, which offer improved approximations of the $\ell_0$-norm relative to that offered by the $\ell_1$-norm [25]. One such alternative leads to solving

$$[\hat{\mathbf{x}}, \hat{\mathbf{o}}] := \arg \min_{\mathbf{x},\mathbf{o}} \frac{1}{2}\|\mathbf{y}-\mathbf{A}\mathbf{x}-\mathbf{o}\|^2_{\mathbf{Q}_w^{-1}} + \lambda_x \sum_{n=1}^{N}\sum_{d=1}^{D_x} \log(|o_{x,n,d}|+\delta_x) + \lambda_y \sum_{n=1}^{N}\sum_{d=1}^{D_y} \log(|o_{y,n,d}|+\delta_y) \quad (17)$$

where $\delta_x$ ($\delta_y$) are small positive constants to ensure that the argument of the logarithm stays away from zero. Since the cost in (17) is non-convex, it is recommended to initialize its iterative minimization with the efficient convex solver of (6). Starting with such an initialization $(\mathbf{x}^{(0)}, \mathbf{o}^{(0)})$, the logarithm can be successively linearized around the $l$-th iterate using $\log(t+\delta) \approx \log(t^{(l)}+\delta)+(t-t^{(l)})/(t^{(l)}+\delta)$ to arrive at a convex cost, which can be readily optimized to obtain the estimates at iteration $(l+1)$. Specifically, at iteration $l$ one solves

$$[\mathbf{x}^{(l)}, \mathbf{o}^{(l)}] := \arg \min_{\mathbf{x},\mathbf{o}} \frac{1}{2}\|\mathbf{y}-\mathbf{A}\mathbf{x}-\mathbf{o}\|^2_{\mathbf{Q}_w^{-1}} + \lambda_x \sum_{n=1}^{N}\sum_{d=1}^{D_x} w^{(l)}_{x,n,d}|o_{x,n,d}| + \lambda_y \sum_{n=1}^{N}\sum_{d=1}^{D_y} w^{(l)}_{y,n,d}|o_{y,n,d}| \quad (18)$$

where

$$w^{(l)}_{x,n,d} := \left(|o^{(l-1)}_{x,n,d}| + \delta_x\right)^{-1}, \quad w^{(l)}_{y,n,d} := \left(|o^{(l-1)}_{y,n,d}| + \delta_y\right)^{-1}.$$

Note that (18) is similar to the DRS one in (6) except that the entries of vector $\mathbf{o}$ in the regularization are weighted non-uniformly. Being convex, (18) can be solved as easily as (6). With reliable initialization offered by the solution of (6), one reason behind the enhancement offered by (18) is the bias correction to Lasso, which is known to yield reliable estimates of the $\mathbf{o}$ support but biased estimates of its nonzero entries [21]. Besides (18), alternative means to mitigate such bias effects is to retain only the outlier-free measurements, after identifying them through the zero entries of $\mathbf{o}$, and use them to re-run the clairvoyant KS which is unbiased. The improvement offered by these refined estimates and those obtained by solving (18) will be corroborated via simulated tests.

## VI. Fixed-lag DRS for online operation

The major limitation of *fixed-interval* smoothing is that the whole batch $\{\mathbf{y}_n\}_{n=1}^N$ has to be available prior to estimating $\{\mathbf{x}_n\}_{n=1}^N$. This is useful for applications such as processing electroencephalograms [36], but not for target tracking. In many tracking applications, state smoothing has to be performed online and stringent delay ("lag") constraints are imposed between the smoothed state instant and the time state estimates are formed. Online smoothing is also important in applications where the state is affected by abrupt changes since these events may be the manifestation of, e.g., system failures [28].

When the outliers are absent in (1), optimal fixed-lag KS can be regarded as a special case of fixed-interval KS [1, p. 176]. The goal here is to estimate $\mathbf{x}_n$, relying upon observations up to time $n+\ell$, where $\ell$ denotes the estimation lag. Supposing that a KF has been run up to time $n$ to yield the state and covariance estimates $\mathbf{x}_{n|n}$ and $\mathbf{\Sigma}_{n|n}$, fixed-lag KS can be accomplished using

$$\begin{aligned}\hat{\mathbf{x}}^{\text{KS}}_{n:n+\ell} = \arg\min_{\mathbf{x}} \quad & \Bigg\{\frac{1}{2}\sum_{n'=n+1}^{n+\ell}\|\mathbf{y}_{n'}-\mathbf{H}_{n'}\mathbf{x}_{n'}\|^2_{\mathbf{R}^{-1}_{n'}} + \frac{1}{2}\|\mathbf{x}_n - \mathbf{x}_{n|n}\|^2_{\mathbf{\Sigma}^{-1}_{n|n}} \\ & + \frac{1}{2}\sum_{n'=n+1}^{n+\ell}\|\mathbf{x}_{n'}-\mathbf{F}_{n'}\mathbf{x}_{n'-1}\|^2_{\mathbf{Q}^{-1}_{n'}}\Bigg\} \end{aligned} \tag{19}$$

where $\hat{\mathbf{x}}^{\text{KS}}_{n:n+\ell} := [(\hat{\mathbf{x}}^{\text{KS}}_n)^T, \ldots, (\hat{\mathbf{x}}^{\text{KS}}_{n+\ell})^T]^T$. Observe that fixed-lag KS in (19) is a special case of fixed-interval KS, when the initial condition on the state, namely $\mathbf{x}_{n|n}$ and $\mathbf{\Sigma}_{n|n}$, is given by the KF, and the state is smoothed over the interval $[n, n+\ell]$. Thus, the solution of (19) can be found with either one of the two algorithms of Section II. However, since (19) does not account for outliers, the resulting estimator is not robust. To address this issue, a fixed-lag DRS is developed next.

## A. Fixed-lag DRS

In the previous section, the fixed-lag KS was regarded as a special case of the fixed-interval KS with properly chosen smoothing interval and initial conditions. Furthermore, in Section III a fixed-interval DRS was developed, which is extended here to robustify the fixed-lag KS in (19). Mimicking the steps followed in Section III to robustify (19) is challenged by the fact that the initial conditions $\mathbf{x}_{n|n}$ and $\mathbf{\Sigma}_{n|n}$ are evaluated by the clairvoyant (and thus non-robust) KF. To overcome this obstacle, the fixed-lag KS will be recast in a form entailing the interval $[n-w, n+\ell]$; that is,

$$\begin{aligned}\check{\mathbf{x}}^{\text{KS}}_{n-w:n+\ell} = \arg\min_{\mathbf{x}} \quad & \Bigg\{\frac{1}{2}\sum_{n'=n-w+1}^{n+\ell}\|\mathbf{y}_{n'}-\mathbf{H}_{n'}\mathbf{x}_{n'}\|^2_{\mathbf{R}^{-1}_{n'}} + \frac{1}{2}\|\mathbf{x}_{n-w} - \mathbf{x}_{n-w|n-w}\|^2_{\mathbf{\Sigma}^{-1}_{n-w|n-w}} \\ & + \frac{1}{2}\sum_{n'=n-w+1}^{n+\ell}\|\mathbf{x}_{n'}-\mathbf{F}_{n'}\mathbf{x}_{n'-1}\|^2_{\mathbf{Q}^{-1}_{n'}}\Bigg\}. \end{aligned} \tag{20}$$

The formulation in (20) is equivalent to that of (19) in the sense that $\check{\mathbf{x}}_n^{\text{KS}} = \hat{\mathbf{x}}_n^{\text{KS}}$. It also suggests that the fixed-lag KS estimate at time $n$ and lag $\ell$ can also be obtained by initializing its recursions with the KF estimates $\mathbf{x}_{n-w|n-w}$ and $\mathbf{\Sigma}_{n-w|n-w}$ for arbitrary $w$. The fixed-lag DRS is obtained by robustifying the fixed-lag KS (20) in a fashion similar to that used in Section III; that is,

$$[\check{\mathbf{x}}_{n-w:n+\ell}^{\text{DRS}}, \check{\mathbf{o}}_{x,n-w:n+\ell}, \check{\mathbf{o}}_{y,n-w:n+\ell}] = \arg\min_{\mathbf{x},\mathbf{o}_x,\mathbf{o}_y} \left\{ \frac{1}{2} \sum_{n'=n-w+1}^{n+\ell} \|\mathbf{y}_{n'} - \mathbf{H}_{n'}\mathbf{x}_{n'} - \mathbf{o}_{y,n'}\|_{\mathbf{R}_{n'}^{-1}}^2 + \frac{1}{2}\|\mathbf{x}_{n-w} - \mathbf{x}_{n-w|n-w}\|_{\mathbf{\Sigma}_{n-w|n-w}^{-1}}^2 + \frac{1}{2} \sum_{n'=n-w+1}^{n+\ell} \|\mathbf{x}_{n'} - \mathbf{F}_{n'}\mathbf{x}_{n'-1} - \mathbf{o}_{x,n'}\|_{\mathbf{Q}_{n'}^{-1}}^2 + \sum_{n'=n-w+1}^{n+\ell} [\lambda_y\|\mathbf{o}_{y,n'}\|_1 + \lambda_x\|\mathbf{o}_{x,n'}\|_1] \right\}. \tag{21}$$

Observe that eventual errors in $\mathbf{x}_{n-w|n-w}$ and $\mathbf{\Sigma}_{n-w|n-w}$ due to the non-robust KF do not severely affect the estimates at time $n$ provided that $w$ is sufficiently large. Certainly, the larger the $w$, the larger the number of nuisance variables involved.

The major limitation of the fixed-lag DRS in (21) is that a convex optimization problem has to be solved at each time $n$ to obtain $\check{\mathbf{x}}_n^{\text{DRS}}$. As a consequence, the associated computational burden to solve the fixed-lag DRS in (21) is not comparable with that of the standard fixed-lag KS. This motivates approximating the fixed-lag approach in (21) to enable online DRS at complexity comparable to that of standard fixed-lag KS and state-of-the-art non-linear smoothers.

### B. Online fixed-lag DRS

The coordinate descent-based fixed-interval algorithm in Section V is properly modified in this section in order to solve the fixed-lag DRS problem formulated in (21). Despite the fact that convergence to a solution of (21) is provably guaranteed asymptotically (i.e., for infinite iterations), satisfactory estimates can be obtained with only a few coordinate descent iterations.

Suppose that between two consecutive observations (say $n+\ell$ and $n+\ell+1$), the affordable delay allows for $J$ coordinate descent iterations to be implemented. Furthermore, for a limited number of iterations, initializing with $\mathbf{o}_{x,n-w:n+\ell}^{(0)}$ and $\mathbf{o}_{y,n-w:n+\ell}^{(0)}$ close to their global optimum values provides a "warm start-up" considerably improving the performance. Observe that for estimating the state at time $n$, fixed-lag DRS entails smoothing over the interval $[n-w, n+\ell]$, and after $J$ coordinate descent iterations, the variables $\mathbf{x}_{n-w:n+\ell}^{(J)}$, $\mathbf{o}_{x,n-w:n+\ell}^{(J)}$, $\mathbf{o}_{y,n-w:n+\ell}^{(J)}$ become available. Since fixed-lag DRS at time $n+1$ entails smoothing over the interval $[n-w+1, n+\ell+1]$, the variables $\mathbf{o}^{(0)}_{x,n-w+1:n+\ell}$, $\mathbf{o}^{(0)}_{y,n-w+1:n+\ell}$ can be initialized to $\mathbf{o}^{(J)}_{x,n-w:n+\ell}$ and $\mathbf{o}^{(J)}_{y,n-w:n+\ell}$ obtained from the previous $J$ iterations, which provides the aforementioned warm start-up. Granted, when the window $w$ is smaller, the effect of the non-robust initialization is more pronounced. Even though no analytical results are claimed on the performance as a function of $w$, the simulated RMSE comparisons in Section VIII with competing alternatives of comparable complexity, speak for the merits of this section's online algorithm.

The novel fixed-lag DRS scheme amounts to sequentially running $J$ KS's and combining their outputs. Interestingly, several non-linear smoothers including those based on SMC and IMM approaches also combine the outputs of several fixed-lag KS's, which allows for a fair comparison of these techniques.

## VII. Generalized linear state-space model

Consider the more general linear state-space model, given by [cf. (1a)]

$$\mathbf{x}_n = \mathbf{F}_n \mathbf{x}_{n-1} + \mathbf{G}_n \mathbf{w}_n + \mathbf{o}_{x,n}, \ \forall n = 1, \ldots, N. \tag{22}$$

where $\{\mathbf{G}_n\}_{n=1}^N$ are known matrices. If matrix $\mathbf{G}_n$ is tall, $\mathbf{G}_n\mathbf{G}_n^T$ is rank deficient, which prevents one from writing the WLS state error as in (4). Instead, KS can be formulated as a constrained optimization problem, and likewise for the corresponding fixed-interval and fixed-lag DRS. Specifically, the novel fixed-interval DRS can be obtained as

$$\begin{aligned}[\hat{\mathbf{x}}^{\text{DRS}}, \hat{\bar{\mathbf{w}}}, \hat{\mathbf{o}}_x, \hat{\mathbf{o}}_y] \quad &:= \quad \arg \min_{\mathbf{x}, \bar{\mathbf{w}}, \mathbf{o}_x, \mathbf{o}_y} C^{\text{DRS}}_{\lambda_x, \lambda_y}(\mathbf{x}, \bar{\mathbf{w}}, \mathbf{o}_x, \mathbf{o}_y) \\ &\qquad \text{subject to} \quad \mathbf{x}_n = \mathbf{F}_n \mathbf{x}_{n-1} + \mathbf{G}_n \mathbf{w}_n + \mathbf{o}_{x,n}, \ \forall \ n = 1, \ldots, N\end{aligned} \tag{23}$$

where $\bar{\mathbf{w}} := [\mathbf{w}_1^T, \mathbf{w}_2^T, \ldots, \mathbf{w}_N^T]^T$ and

$$\begin{aligned}C^{\text{DRS}}_{\lambda_x, \lambda_y}(\mathbf{x}, \bar{\mathbf{w}}, \mathbf{o}_x, \mathbf{o}_y) \quad = \quad & \frac{1}{2}\sum_{n=1}^{N} \|\mathbf{y}_n - \mathbf{H}_n \mathbf{x}_n - \mathbf{o}_{y,n}\|^2_{\mathbf{R}_n^{-1}} + \frac{1}{2}\|\mathbf{x}_0 - \mathbf{m}_0\|^2_{\mathbf{\Sigma}_0^{-1}} + \frac{1}{2}\sum_{n=1}^{N} \|\mathbf{w}_n\|^2_{\mathbf{Q}_n^{-1}} \\ & + \sum_{n=1}^{N} [\lambda_x \|\mathbf{o}_{x,n}\|_1 + \lambda_y \|\mathbf{o}_{y,n}\|_1].\end{aligned}$$

Due to the constrained nature of the problem in (23), coordinate descent iterations can not be directly applied. However, it is possible to develop iterations based on the alternating direction method of multipliers (AD-MoM) [7]. These iterations are simple if one introduces the auxiliary variables $\mathbf{a}_n = \mathbf{o}_{x,n}$ and $\mathbf{b}_n = \mathbf{o}_{y,n}$ which imply additional constraints. Then, the augmented Lagrangian can be written as

$$\begin{aligned}
\mathcal{L}_\kappa &= \frac{1}{2}\sum_{n=1}^{N}\|\mathbf{y}_n - \mathbf{H}_n\mathbf{x}_n - \mathbf{o}_{y,n}\|^2_{\mathbf{R}_n^{-1}} + \frac{1}{2}\|\mathbf{x}_0 - \mathbf{m}_0\|^2_{\boldsymbol{\Sigma}_0^{-1}} + \frac{1}{2}\sum_{n=1}^{N}\|\mathbf{w}_n\|^2_{\mathbf{Q}_n^{-1}} + \sum_{n=1}^{N}[\lambda_y\|\mathbf{b}_n\|_1 + \lambda_x\|\mathbf{a}_n\|_1] \\
&+ \sum_{n=1}^{N}\left[\boldsymbol{\chi}_n^T(\mathbf{x}_n - \mathbf{F}_n\mathbf{x}_{n-1} - \mathbf{G}_n\mathbf{w}_n - \mathbf{o}_{x,n}) + \frac{\kappa}{2}\|\mathbf{x}_n - \mathbf{F}_n\mathbf{x}_{n-1} - \mathbf{G}_n\mathbf{w}_n - \mathbf{o}_{x,n}\|_2^2\right] \\
&+ \sum_{n=1}^{N}\left[\boldsymbol{\mu}_n^T(\mathbf{o}_{y,n} - \mathbf{b}_n) + \frac{\kappa}{2}\|\mathbf{o}_{y,n} - \mathbf{b}_n\|_2^2\right] + \sum_{n=1}^{N}\left[\boldsymbol{\nu}_n^T(\mathbf{o}_{x,n} - \mathbf{a}_n) + \frac{\kappa}{2}\|\mathbf{o}_{x,n} - \mathbf{a}_n\|_2^2\right]. \qquad (24)
\end{aligned}$$

where $\{\boldsymbol{\chi}_n,\ \boldsymbol{\mu}_n,\ \boldsymbol{\nu}_n\}_{n=1}^N$ denote the Lagrange multipliers and $\kappa$ is a positive constant. Setting the derivatives of $\mathcal{L}_\kappa$ with respect to $\mathbf{x}_n$ equal to zero, yields the following AD-MoM iteration [cf. (24)]

$$\begin{aligned}
\mathbf{x}^{(j)} &= \arg\min_{\mathbf{x}}\left\{\frac{1}{2}\sum_{n=1}^{N}\|\mathbf{y}_n - \mathbf{H}_n\mathbf{x}_n - \mathbf{o}_{y,n}^{(j-1)}\|^2_{\mathbf{R}_n^{-1}} + \frac{1}{2}\|\mathbf{x}_0 - \mathbf{m}_0\|^2_{\boldsymbol{\Sigma}_0^{-1}}\right. \\
&\left. + \sum_{n=1}^{N}\frac{\kappa}{2}\|\mathbf{x}_n - \mathbf{F}_n\mathbf{x}_{n-1} - \mathbf{G}_n\mathbf{w}_n^{(j-1)} - \mathbf{o}_{x,n}^{(j-1)} + \boldsymbol{\chi}_n^{(j-1)^T}/\kappa\|_2^2\right\}.
\end{aligned}$$

Clearly, this problem is equivalent to (4), which can be solved in a batch or recursive form at complexity that is linear in $N$. Likewise, the remaining variables are updated as follows:

$$\begin{aligned}
\mathbf{w}_n^{(j)} &= (\mathbf{Q}_n^{-1} + \kappa\mathbf{G}_n^T\mathbf{G}_n)^{-1}\mathbf{G}_n^T(\boldsymbol{\chi}_n^{(j-1)} + \kappa\mathbf{x}_n^{(j)} - \kappa\mathbf{F}_n\mathbf{x}_{n-1}^{(j)} - \kappa\mathbf{o}_{x,n}^{(j-1)}),\ n = 1,\ldots,N \\
\mathbf{o}_{y,n}^{(j)} &= (\mathbf{R}_n^{-1} + \kappa\mathbf{I}_{D_y})^{-1}\big(\mathbf{R}_n^{-1}(\mathbf{y}_n - \mathbf{H}_n\mathbf{x}_n^{(j)}) - \boldsymbol{\mu}_n^{(j-1)^T} + \kappa\mathbf{b}_n^{(j-1)}\big),\ n = 1,\ldots,N \\
\mathbf{o}_{x,n}^{(j)} &= \frac{1}{2}\big(\boldsymbol{\chi}_n^{(j-1)}/\kappa + \mathbf{x}_n^{(j)} - \mathbf{F}_n\mathbf{x}_{n-1}^{(j)} - \mathbf{G}_n\mathbf{w}_n^{(j)} + \mathbf{a}_n^{(j-1)} - \boldsymbol{\nu}_n^{(j-1)}/\kappa\big),\ n = 1,\ldots,N \\
b_{n,d}^{(j)} &= \frac{1}{\kappa}\max\big(|\kappa o_{y,n,d}^{(j)} + \mu_{n,d}^{(j-1)}| - \lambda_y, 0\big)\mathrm{sign}\big(\kappa o_{y,n,d}^{(j)} + \mu_{n,d}^{(j-1)}\big),\ n = 1,\ldots,N,\ d = 1,\ldots,D_y \\
a_{n,d}^{(j)} &= \frac{1}{\kappa}\max\big(|\kappa o_{x,n,d}^{(j)} + \nu_{n,d}^{(j-1)}| - \lambda_x, 0\big)\mathrm{sign}\big(\kappa o_{x,n,d}^{(j)} + \nu_{n,d}^{(j-1)}\big),\ n = 1,\ldots,N,\ d = 1,\ldots,D_x \\
\boldsymbol{\chi}_n^{(j)} &= \boldsymbol{\chi}_n^{(j-1)} + \kappa(\mathbf{x}_n^{(j)} - \mathbf{F}_n\mathbf{x}_{n-1}^{(j)} - \mathbf{G}_n\mathbf{w}_n^{(j)} - \mathbf{o}_{x,n}^{(j)}),\quad n = 1,\ldots,N \\
\boldsymbol{\mu}_n^{(j)} &= \boldsymbol{\mu}_n^{(j-1)} + \kappa(\mathbf{o}_{y,n}^{(j)} - \mathbf{b}_n^{(j)}),\quad n = 1,\ldots,N \\
\boldsymbol{\nu}_n^{(j)} &= \boldsymbol{\nu}_n^{(j-1)} + \kappa(\mathbf{o}_{x,n}^{(j)} - \mathbf{a}_n^{(j)}),\quad n = 1,\ldots,N.
\end{aligned}$$

Invoking the results in [7, p. 256], guarantees global convergence of these iterations as asserted next.

**Proposition 5.** *For any $\kappa > 0$ and arbitrary initial values $\mathbf{w}^{(0)}, \mathbf{o}_y^{(0)}, \mathbf{o}_x^{(0)}, \mathbf{b}^{(0)}, \mathbf{a}^{(0)}, \boldsymbol{\chi}^{(0)}, \boldsymbol{\mu}^{(0)}, \boldsymbol{\nu}^{(0)}$, the AD-MoM iterates are all convergent. Furthermore, every limit point of the sequences $\mathbf{x}^{(j)}$, $\mathbf{w}^{(j)}$, $\mathbf{o}_y^{(j)}$, $\mathbf{o}_x^{(j)}$ is a solution of the problem in* (23).

Although global convergence of the coordinate descent and AD-MoM iterates is ensured by Propositions 4 and 5, respectively, no analytical results are available in optimization theory regarding their rate of convergence – a challenging subject going well beyond the scope of the present work.

For AD-MoM iterations too, the bulk of computations is in the order of a KS, which grows linearly in $N$. The rest involves closed-form evaluations. In a nutshell, for the general linear model in (22) the AD-MoM iterations replace those of the coordinate descent algorithm with the same order of computational complexity. Some of the simulations in the ensuing section will test this AD-MoM based fixed-interval DRS approach, which also has a fixed-lag counterpart tailored for online operation under the general linear state-space model. Its derivation follows closely that of the coordinate descent for fixed-interval DRS, and is omitted for brevity.

## VIII. Simulated tests: Maneuvering target tracking with glint

In this section, the developed robust smoothers are simulated for maneuvering target tracking in the presence of glint noise. First, DRS performance is tested on a sample target trajectory, and then sample averaged performance metrics for DRS are compared against the main competing alternatives.

### A. DRS on a Sample Trajectory

The model in (22) is simulated with $\mathbf{x}_n := [p_n^x, s_n^x, p_n^y, s_n^y]^T$, where $p_n^x$ and $p_n^y$ denote the target position in the $x$ and $y$ coordinates, respectively; and correspondingly $s_n^x$ and $s_n^y$ denote the target velocity in the $x$ and $y$ directions; thus, $D_x = 4$. The matrices in (1b) and (22) are invariant $\forall\ n$

$$\mathbf{F}_n := \begin{pmatrix} 1 & \tau & 0 & 0 \\ 0 & 1 & 0 & 0 \\ 0 & 0 & 1 & \tau \\ 0 & 0 & 0 & 1 \end{pmatrix}, \quad \mathbf{G}_n := \begin{pmatrix} \frac{\tau^2}{2} & 0 \\ \tau & 0 \\ 0 & \frac{\tau^2}{2} \\ 0 & \tau \end{pmatrix}, \quad \mathbf{H}_n := \begin{pmatrix} 1 & 0 & 0 & 0 \\ 0 & 0 & 1 & 0 \end{pmatrix} \tag{25}$$

and $\tau$ denotes the sampling period. Since $\mathbf{G}_n$ is tall, this so-termed discrete white noise acceleration (DWNA) model [4, p. 273], can only be handled by the generalized linear state-space model of Section VII. The form of $\mathbf{H}_n$ in (25), shows that $\mathbf{y}_n$ comprises noisy position measurements, and $D_y = 2$.

A total of $N = 100$ observations are collected, $\tau = 1$, $\mathbf{R}_n = 150^2\mathbf{I}_2$, $\mathbf{Q}_n = 0.5\mathbf{I}_2$, $\mathbf{m}_0 = \mathbf{0}_4$, and $\mathbf{\Sigma}_0 = \text{diag}(50, 5, 50, 5)$. The target trajectory starts from position $[0, 0]$, and evolves according to the DWNA model with the specified parameters from time $n = 1$ to $n = 29$. At times $n = 30$ and 31 the target turns right and follows again the DWNA model from $n = 32$ to 59. At time $n = 60$ and 61 the target turns left and then proceeds with the DWNA model until $n = N$. The true target trajectory is depicted in Fig. 2 with solid line. The circles represent the acquired position measurements. Three outliers (not depicted in the figure) yield erroneous position reports, at $n = 15$, 50, and 80.

Figure 3 depicts the clairvoyant fixed-interval KS estimate. Observe that KS is not robust to outliers in the observations and state. The fixed-interval DRS estimates shown in Fig. 4 for $\lambda_y = 0.01$, and $\lambda_x = 0.05$ (which approximately correspond to 10% of the critical $\bar{\lambda}_y$ and $\bar{\lambda}_x$ in Proposition 3), demonstrate that DRS can effectively cope with outliers, and has merits over the non-robust KS even when $(\lambda_x, \lambda_y)$ are not systematically estimated as in (9) or (10). Figures 5 and 6 depict estimates of the fixed-lag KS in (19), and DRS in (21) for lag $\ell = 10$, respectively. Again, the KS estimates are strongly affected by outliers. On the other hand, the fixed-lag DRS estimates in Fig. 6, for $w = \ell$, $\lambda_y = 0.01$, and $\lambda_x = 0.05$, are only minimally affected by outliers.

### B. *Online Fixed-lag DRS versus Rao-Blackwellized SMC*

The root mean-square error (RMSE) of the position estimates is used here to quantify the performance improvement of DRS relative to KS. The true target trajectory coincides with that of Fig. 2 (solid line), when $M = 100$ noise and outlier realizations are present. With probability $\pi = 0.97$, the model in (1b) was in effect with $\mathbf{o}_y = \mathbf{0}$, $\mathbf{H}_n$ in (25), and $\mathbf{R}_n = 150^2\mathbf{I}_2$. With probability $1 - \pi = 0.03$, outliers in the observations occur, and in this case the position reports are $[y_{n,1}, y_{n,2}] \sim \mathcal{U}([-10000, 10000]^2)$. Figure 7 depicts the RMSE of the position estimates, $\text{RMSE}_n = \sqrt{\frac{1}{M}\sum_{m=1}^{M}[(p_n^x - \widehat{p}_n^{x(m)})^2 + (p_n^y - \widehat{p}_n^{y(m)})^2]}$, where $[\widehat{p}_n^{x(m)}, \widehat{p}_n^{y(m)}]$ is the estimated position at time $n$ for the $m$th noise and outlier realization, for the fixed-interval KS and DRS with $\lambda_y = 0.01$, and $\lambda_x = 0.05$. Clearly, DRS exhibits lower RMSE than the clairvoyant KS.

Figure 8 depicts the instantaneous RMSE for fixed-lag KS and DRS for $\ell = 10$, $w = 10$, $\lambda_y = 0.01$, and $\lambda_x = 0.05$, along with the Rao-Blackwellized (RB) SMC smoother relying on 50 particles, and the online fixed-lag DRS with 50 AD-MoM iterations and constant $\kappa = 0.05$ (referred in the figure as O-DRS). For the RB-SMC smoother, a conditionally linear, Gaussian model is adopted. Specifically, under nominal conditions, the model is that in (1) with $\mathbf{F}_n$, $\mathbf{G}_n$, and $\mathbf{H}_n$ as in (25), $\mathbf{o} = \mathbf{0}$, $\mathbf{Q}_n = 0.5\mathbf{I}_2$, and $\mathbf{R}_n = 150^2\mathbf{I}_2$. It is assumed that the nominal conditions are in effect with probability $\pi = 0.97$. Outliers in the measurements occur with probability $1-\pi = 0.03$, and in this case $\mathbf{R}_n = 15000^2\mathbf{I}_2$, which allows for down-weighting the respective measurements. With the same probability, state outliers emerge too, and in this case $\mathbf{Q}_n = 500\mathbf{I}_2$. Clearly, conditioned on the outlier realizations, the dynamical process is linear and Gaussian. This allows for drawing particles for the state/measurement outliers, and using them for estimating the state via fixed-lag KS; see also [13] for details on the fixed-lag RB-SMC. In addition to fixed-lag KS, the novel O-DRS approach outperforms RB-SMC for the same computational burden.

### *C. Comparison with RANSAC and Huber M-estimates*

DRS is compared here against state-of-the-art robust smoothers, namely the Huber based scheme, and a combined RANSAC followed by Huber scheme. In the latter, smoothing is cast as the linear regression problem in (3) to which RANSAC can be applied readily [20]. RANSAC relies on random draws (here 100 or 1,000) to find the "best" possible subset of rows corresponding to the nominal model [14], [20]. To ensure that the remaining outliers do not degrade performance, the nominal rows of (3) found by RANSAC are pre-whitened, and subsequently plugged into Huber's cost in (7). The Huber parameters are set to $\lambda_x = \lambda_y = 1.345$ as suggested by [17], [23] for standardized Gaussian nominal noise (this requires pre-whitening the nominal noise). The Huber estimate is found by solving (7) using the iteratively re-weighted least-squares (IRLS) algorithm in [23], which unlike the Lasso-based solver pursued here, guarantees only local convergence. The fixed-interval DRS in (6) is also employed with $\lambda_x$ and $\lambda_y$ found using either of the two data-driven criteria suggested in Section IV. To further improve DRS, one iteration of the refined estimate in Remark 3 is also implemented. The model simulated here obeys (25), but with $\mathbf{G}_n = \mathbf{I}_4$, $\mathbf{w}_n \sim \mathcal{N}(\mathbf{0}_4, \mathbf{Q}_n)$, $\mathbf{Q}_n = \text{diag}(1, 0.001, 1, 0.001)$, and $\mathbf{R}_n = 5\mathbf{I}_2$. State and measurement outliers are generated as independent Laplacian with variances 200 and 20,000, respectively. The RMSE for both position and velocity estimates time-averaged over 100 Monte Carlo runs is plotted versus the percentage of outlier contamination.

Fig. 9 plots the RMSE versus percentage of outliers for the combined RANSAC-Huber robust smoother as well as DRS, when outliers appear only in the measurements. The numerical suffix for DRS denotes the grid size used for the AVD [cf. (10)], while the one for RANSAC stands for the number of RANSAC's random draws. In terms of complexity, DRS-100 (with $I_x = I_y = 10$ grid points equispaced in log scale as suggested in [15]) lies in-between RANSAC-100 and RANSAC-1000. Note that up to $50\%$ outliers all three methods perform similarly. When the outlier contamination percentage exceeds $50\%$, RANSAC-100 performs poorly, while RANSAC-1000 and DRS-100 exhibit graceful performance degradation. Due to its lower-complexity, DRS-100 offers a better alternative than RANSAC-1000.

Fig. 10 compares RMSE performance when outliers are only present in the state. One observes that DRS-100 considerably outperforms both versions of RANSAC for all percentages of outlier contamination. The improvement in going from RANSAC-100 to RANSAC-1000 is not noticeable.

Fig. 11 plots the RMSE resulting from DRS-100, batch KS, Huber-only, and the combined RANSAC-Huber scheme when outliers are simultaneously present in the state and measurements. The AVD criterion is used for DRS. It can be seen that RANSAC-Huber combination performs poorly for outliers present in the state and measurement. This happens because RANSAC removes certain rows of the regression matrix, namely those contaminated by outliers, which renders the remaining sub-matrix ill-conditioned. Huber-only performs close but worse than KS – a manifestation of the fact that Huber's estimate are found for independent nominal noise. Even though the noise here is independent, it is not standard Gaussian and the subsequent pre-whitening, which is a mere scaling in this case, adversely affects the Huber-based estimate. Indeed, neither of the mentioned robust methods performs well when outliers are present both in the state and measurement, and surprisingly even the clairvoyant KS outperforms them. However, DRS-100 outperforms KS in terms of RMSE, and speaks for the importance of the universality property of the novel estimator.

The DRS improvement over KS is more pronounced if the percentage of outliers is known, case where (9) is used instead of AVD. The result is plotted in Fig. 12, where DRS significantly outperforms KS. Here, the percentage of state outliers is fixed at $10\%$, while that of measurement outliers is variable.

At last, different DRS renditions are compared against each other and with the robust smoother of [3]. For a fair comparison with [3], the setup includes outliers only in the measurements and smoothed estimates for both approaches are formed using the general-purpose optimization software SeDuMi [34]. Each randomly occurring outlying-measurement is drawn from a zero-mean uniform distribution with variance $20,000$ independently from the nominal random variables. Note that the outlying measurements here are generated not in accordance with the model in [3] in order to illustrate the universality attribute of the proposed DRS. Nominal model parameters commonly known to both DRS and the smoother [3], are chosen as before with the only difference that $\mathbf{Q}_n = \text{diag}(100, 1, 100, 1)$. Fig. 13 depicts the mean and median RMSE computed over $1,000$ Monte Carlo runs as a function of the percentage of outliers. It can be seen that DRS with AVD outperforms the smoother in [3], especially as the fraction of outlier contamination exceeds 10%. Similar to the smoother in [3], DRS with AVD utilizes only nominal parameter knowledge to form its estimate. The curve that utilizes the concave penalty pertains to the refinement outlined in Remark 3. A clear gain is observed with this refinement as a result of de-biasing the DRS estimates. DRS with known percentage of outliers is also plotted and outperforms all other alternatives. This is due to the extra knowledge on outlier sparsity that this smoother benefits from.

## IX. CONCLUSIONS

Robust smoothers were developed for dynamical processes contaminated with outliers in the observations and/or state. The novel fixed-interval DRS can be viewed as an $\ell_1$-norm regularized version of the WLS-based clairvoyant KS algorithm. This form of regularization controls the sparsity of outliers, which are explicitly introduced as auxiliary variables. Two data-driven methods were also devised to select the associated regularization parameters. Block coordinate descent-based iterations were developed to solve the underlying convex optimization problem in an efficient manner. To enable real-time smoothing for delay-constrained applications such as target tracking, an online fixed-lag DRS was also developed. At last, the novel approach was broadened to include generalized linear state-space models. Numerical tests demonstrated that the proposed algorithms can jointly cope with state and measurement outliers, and outperform state-of-the-art methods at comparable computational burden.

## APPENDIX

### A. *Proof of Proposition 1 (MAP optimality of DRS in* (6)*)*

Successive application of Bayes' rule, as well as the assumptions on independence and the corresponding distributions of the nominal noise and outlier vectors yield

$$p(\mathbf{x}, \mathbf{o}_x, \mathbf{o}_y | \mathbf{y}_{1:N}) = \frac{p(\mathbf{y}_{1:N}, \mathbf{x}, \mathbf{o}_x, \mathbf{o}_y)}{p(\mathbf{y}_{1:N})} \propto \; p(\mathbf{x}_0) \prod_{n=1}^{N} p(\mathbf{y}_n | \mathbf{x}_{0:n}, \mathbf{y}_{1:n-1}, \mathbf{o}_{x,1:n}, \mathbf{o}_{y,1:n})$$
$$\times p(\mathbf{x}_n | \mathbf{x}_{0:n-1}, \mathbf{y}_{1:n-1}, \mathbf{o}_{x,1:n}, \mathbf{o}_{y,1:n}) p(\mathbf{o}_{y,n} | \mathbf{x}_{0:n-1}, \mathbf{y}_{1:n-1}, \mathbf{o}_{x,1:n}, \mathbf{o}_{y,1:n-1})$$
$$\times p(\mathbf{o}_{x,n} | \mathbf{x}_{0:n-1}, \mathbf{y}_{1:n-1}, \mathbf{o}_{x,1:n-1}, \mathbf{o}_{y,1:n-1}) = p(\mathbf{x}_0) \prod_{n=1}^{N} p(\mathbf{y}_n | \mathbf{x}_n, \mathbf{o}_{y,n}) p(\mathbf{x}_n | \mathbf{x}_{n-1}, \mathbf{o}_{x,n}) p(\mathbf{o}_{y,n}) p(\mathbf{o}_{x,n})$$

$$= \mathcal{N}(\mathbf{x}_0; \mathbf{m}_0, \mathbf{\Sigma}_0) \prod_{n=1}^{N} \mathcal{N}(\mathbf{y}_n; \mathbf{H}_n\mathbf{x}_n + \mathbf{o}_{y,n}, \mathbf{R}_n)\mathcal{N}(\mathbf{x}_n; \mathbf{F}_n\mathbf{x}_{n-1} + \mathbf{o}_{x,n}, \mathbf{Q}_n)\mathcal{L}(\mathbf{o}_{x,n}; \lambda_x)\mathcal{L}(\mathbf{o}_{y,n}; \lambda_y) \quad (26)$$

where $\mathcal{N}(\mathbf{x}; \mathbf{m}_0, \mathbf{\Sigma}_0)$ represents a Gaussian distribution with mean $\mathbf{m}_0$ and covariance $\mathbf{\Sigma}_0$, while $\mathcal{L}(\mathbf{o}; \lambda) := \prod_d (\lambda/2) \exp(-\lambda|o_d|) \propto \exp(-\lambda\|\mathbf{o}\|_1)$ represents the joint Laplacian distribution for a vector with independent entries. Maximizing (26) amounts to minimizing the negative of the exponent, which leads to the DRS criterion in (6).

### B. *Proof of Proposition 2 (Equivalence of* (6) *with* (7)*)*

Consider minimizing the cost in (6) over $\mathbf{o}_y$ and $\mathbf{o}_x$, with $\mathbf{x}$ fixed. Given $\mathbf{x}$ and with $\{\mathbf{Q}_n, \mathbf{R}_n\}_{n=1}^N$ and $\mathbf{\Sigma}_0$ given by identity matrices, the criterion in (6) is separable over each scalar entry of $\mathbf{o}_y$ and $\mathbf{o}_x$. Hence, it suffices to find

$$\widehat{o}_{y,n,d} = \arg\min_{o_{y,n,d}} \left\{ \frac{1}{2}(y_{n,d} - \mathbf{h}_{n,d}^T\mathbf{x}_n - o_{y,n,d})^2 + \lambda_y|o_{y,n,d}| \right\}, n = 1, \ldots, N, \; d = 1, \ldots, D_y \quad (27a)$$

$$\widehat{o}_{x,n,d} = \arg\min_{o_{x,n,d}} \left\{ \frac{1}{2}(x_{n,d} - \mathbf{f}_{n,d}^T\mathbf{x}_{n-1} - o_{x,n,d})^2 + \lambda_x|o_{x,n,d}| \right\}, n = 1, \ldots, N, \; d = 1, \ldots, D_x \quad (27b)$$

The scalar problems in (27) admit, respectively, the following closed-form solutions (see, e.g., [29]):

$$\widehat{o}_{y,n,d} = \begin{cases} 0, & \text{if } |y_{n,d} - \mathbf{h}_{n,d}^T\mathbf{x}_n| \leq \lambda_y \\ y_{n,d} - \mathbf{h}_{n,d}^T\mathbf{x}_n - \lambda_y \text{ sign}(y_{n,d} - \mathbf{h}_{n,d}^T\mathbf{x}_n), & \text{otherwise} \end{cases} \quad (28a)$$

$$\widehat{o}_{x,n,d} = \begin{cases} 0, & \text{if } |x_{n,d} - \mathbf{f}_{n,d}^T\mathbf{x}_{n-1}| \leq \lambda_x \\ x_{n,d} - \mathbf{f}_{n,d}^T\mathbf{x}_{n-1} - \lambda_x \text{ sign}(x_{n,d} - \mathbf{f}_{n,d}^T\mathbf{x}_{n-1}), & \text{otherwise} \end{cases} \quad (28b)$$

Substituting (28) in the DRS cost of (6), the subsequent optimization problem in $\mathbf{x}$ is

$$\begin{aligned} \widehat{\mathbf{x}} := \arg\min_{\mathbf{x}} \Bigg[ & \sum_{n=1}^{N}\sum_{d=1}^{D_y} \Bigg( \frac{1}{2}(y_{n,d} - \mathbf{h}_{n,d}^T\mathbf{x}_n)^2 \mathbb{1}_{|y_{n,d} - \mathbf{h}_{n,d}^T\mathbf{x}_n| \leq \lambda_y}(\mathbf{x}) \\ & + \big(\lambda_y|y_{n,d} - \mathbf{h}_{n,d}^T\mathbf{x}_n| - \lambda_y^2/2\big) \mathbb{1}_{|y_{n,d} - \mathbf{h}_{n,d}^T\mathbf{x}_n| > \lambda_y}(\mathbf{x}) \Bigg) + \sum_{d=1}^{D_x} \left( \frac{1}{2}(x_{0,d} - m_{0,d})^2 \right) \\ & + \sum_{n=1}^{N}\sum_{d=1}^{D_x} \Bigg( \frac{1}{2}(x_{n,d} - \mathbf{f}_{n,d}^T\mathbf{x}_{n-1})^2 \mathbb{1}_{|x_{n,d} - \mathbf{f}_{n,d}^T\mathbf{x}_{n-1}| \leq \lambda_x}(\mathbf{x})) \\ & + \big(\lambda_x|x_{n,d} - \mathbf{f}_{n,d}^T\mathbf{x}_{n-1}| - \lambda_x^2/2\big) \mathbb{1}_{|x_{n,d} - \mathbf{f}_{n,d}^T\mathbf{x}_{n-1}| > \lambda_x}(\mathbf{x}) \Bigg) \Bigg]. \end{aligned} \quad (29)$$

Given the definition of Huber's cost, the problem in (29) is equivalent to (7). Therefore, (6) and (7) are equivalent.

### C. Proof of Proposition 3

Letting $L(\mathbf{x}, \mathbf{o}_x, \mathbf{o}_y)$ denote the cost in (6) and $\check{\nabla}$ the subgradient operator, the optimality conditions for the non-differentiable problem in (6) are [32, p. 126]

$$\mathbf{0} \in \check{\nabla}_{\mathbf{o}_{y,n}} L(\hat{\mathbf{x}}, \hat{\mathbf{o}}_x, \hat{\mathbf{o}}_y) \;\; \Rightarrow \;\; \mathbf{0} \in \mathbf{R}_n^{-1}\hat{\mathbf{o}}_{y,n} - \mathbf{R}_n^{-1}(\mathbf{y}_n - \mathbf{H}_n\hat{\mathbf{x}}_n) + \lambda_y \check{\mathbf{o}}_{y,n} \tag{30a}$$

$$\mathbf{0} \in \check{\nabla}_{\mathbf{o}_{x,n}} L(\hat{\mathbf{x}}, \hat{\mathbf{o}}_x, \hat{\mathbf{o}}_y) \;\; \Rightarrow \;\; \mathbf{0} \in \mathbf{Q}_n^{-1}\hat{\mathbf{o}}_{x,n} - \mathbf{Q}_n^{-1}(\hat{\mathbf{x}}_n - \mathbf{F}_n\hat{\mathbf{x}}_{n-1}) + \lambda_x \check{\mathbf{o}}_{x,n} \tag{30b}$$

where $\check{\mathbf{o}}_{x,n} := [\check{o}_{x,n,1}, \check{o}_{x,n,2}, \ldots, \check{o}_{x,n,D_x}]^T$ and $\check{\mathbf{o}}_{y,n} := [\check{o}_{y,n,1}, \check{o}_{y,n,2}, \ldots, \check{o}_{y,n,D_y}]^T$ are the subgradients of $\|\mathbf{o}_{x,n}\|_1$ and $\|\mathbf{o}_{y,n}\|_1$, respectively, whose $d$th entries are given by

$$\check{o}_{y,n,d} = \begin{cases} \text{sign}(\hat{o}_{y,n,d}), & \hat{o}_{y,n,d} \neq 0 \\ s_{n,d}, & \hat{o}_{y,n,d} = 0 \end{cases}, \quad \check{o}_{x,n,d} = \begin{cases} \text{sign}(\hat{o}_{x,n,d}), & \hat{o}_{x,n,d} \neq 0 \\ t_{n,d}, & \hat{o}_{x,n,d} = 0 \end{cases},$$

for any $|s_{n,d}| \leq 1$ and $|t_{n,d}| \leq 1$.

DRS coincides with KS when $\hat{\mathbf{o}}_y = \mathbf{0}_{ND_y}$ and $\hat{\mathbf{o}}_x = \mathbf{0}_{ND_x}$, which implies $\hat{\mathbf{x}} := \hat{\mathbf{x}}^{\text{KS}}$. For $\hat{\mathbf{o}}_y = \mathbf{0}_{ND_y}$, (30a) is satisfied with $\hat{\mathbf{x}} := \hat{\mathbf{x}}^{\text{KS}}$ if and only if (8a) holds. Similarly, for $\hat{\mathbf{o}}_x = \mathbf{0}_{ND_x}$, (30b) is satisfied with $\hat{\mathbf{x}} := \hat{\mathbf{x}}^{\text{KS}}$ if and only if (8b) holds. QED

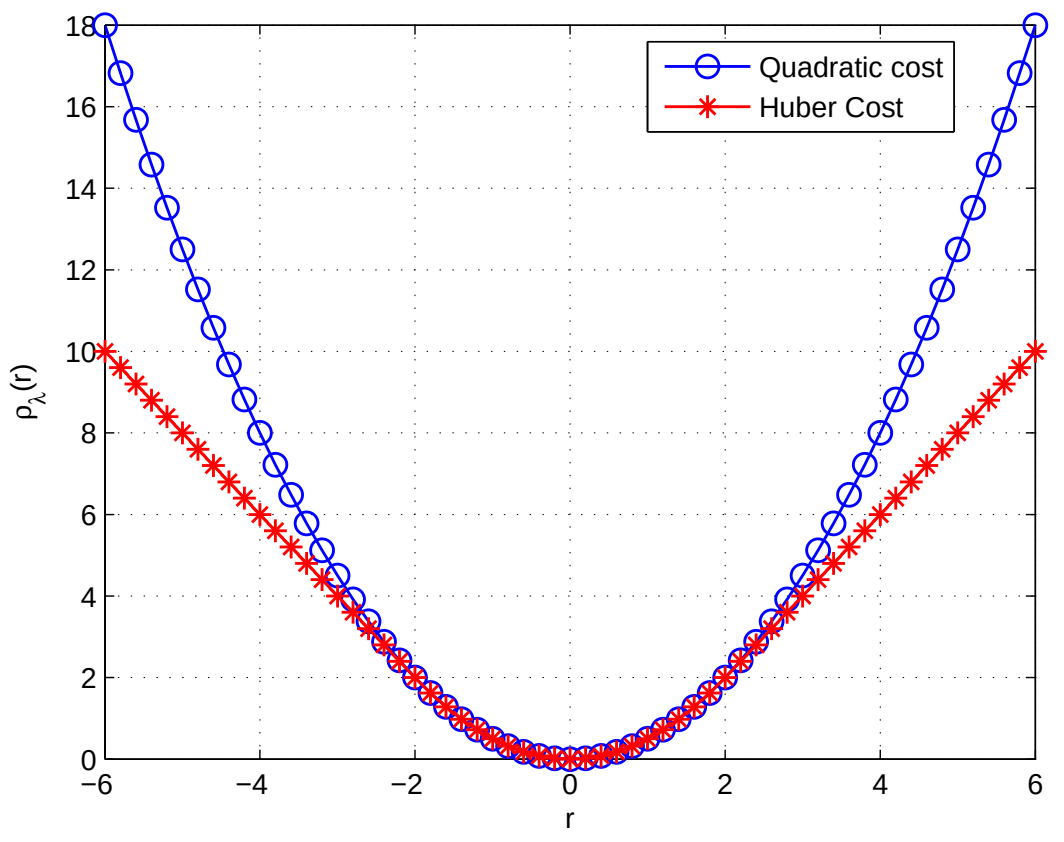

Fig. 1: Quadratic cost versus Huber cost ($\lambda = 2$).

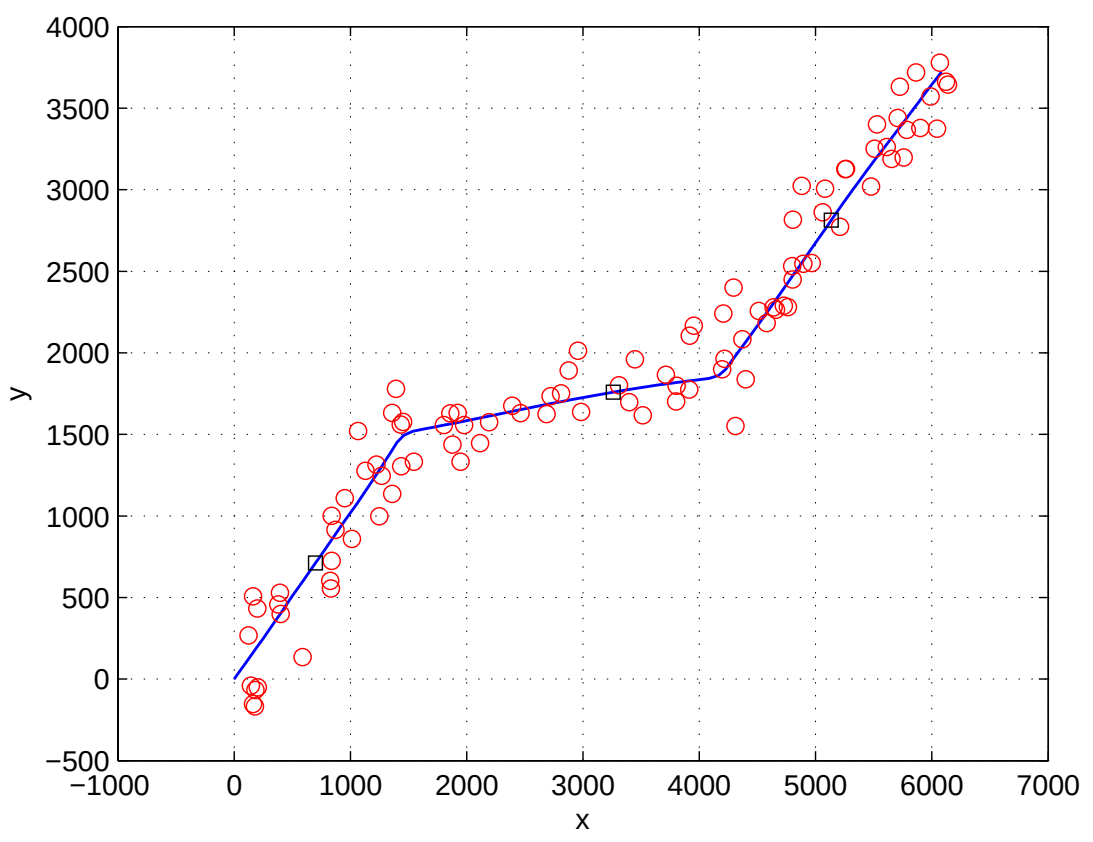

Fig. 2: True target trajectory (solid line); Observed positions (circles). The squares indicate the trajectory instants where outliers occur ($n = 15, 50$, and $80$). Outlier-corrupted measurement values are $\mathbf{y}_{15} = [-5560, 18440]^T$, $\mathbf{y}_{50} = [3880, 14440]^T$, and $\mathbf{y}_{80} = [6440, -14800]^T$.

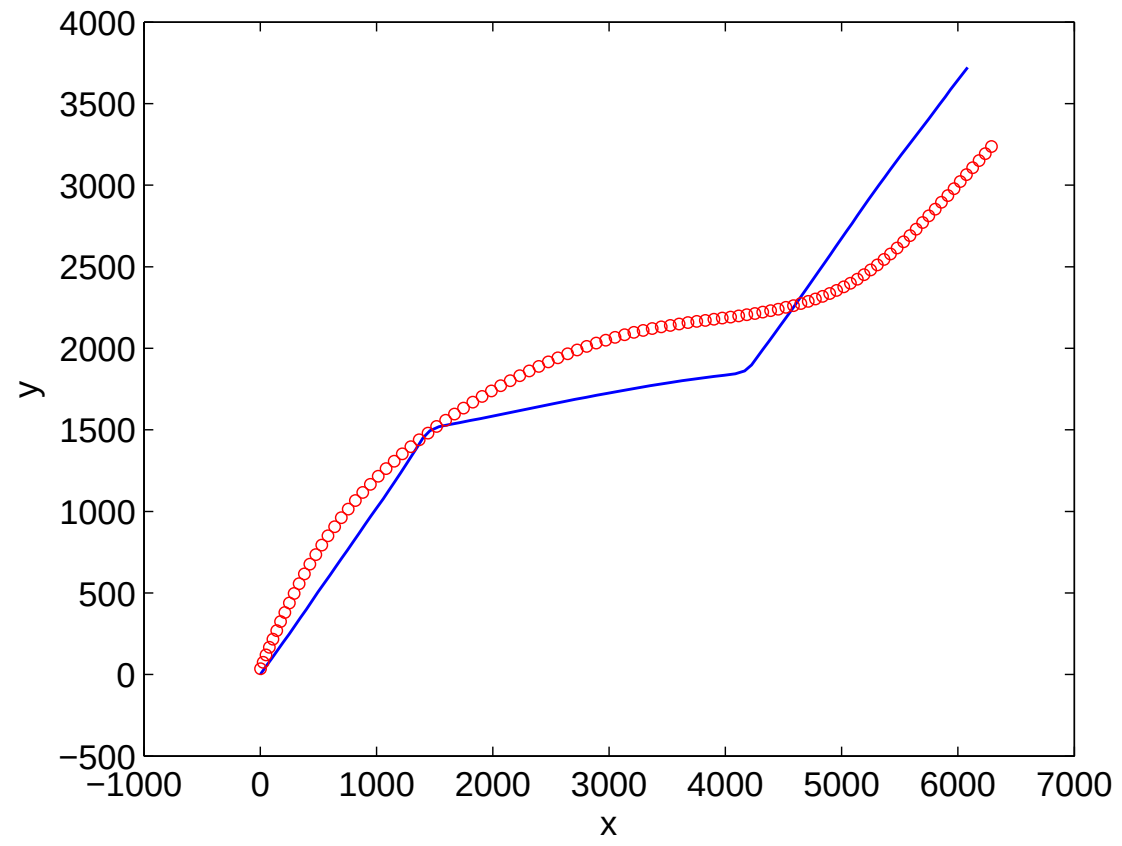

Fig. 3: True target trajectory (solid line) and estimated trajectory (circles) using fixed-interval KS.

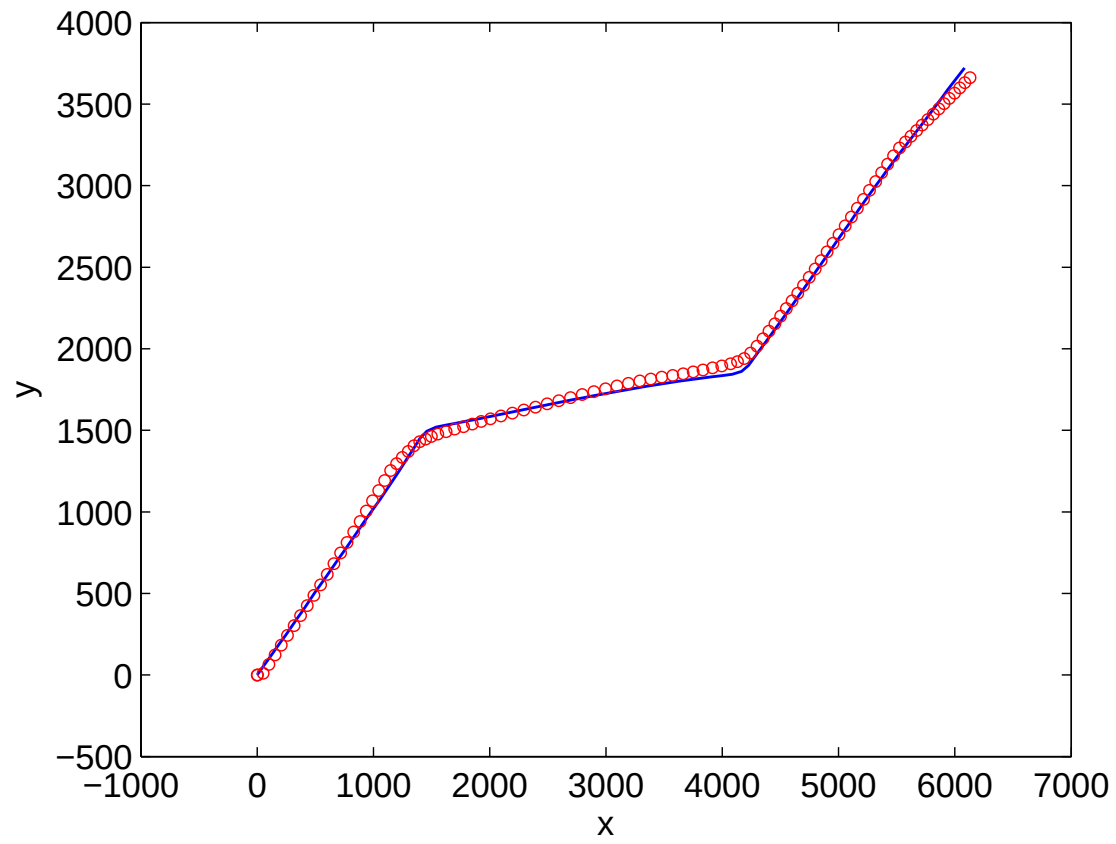

Fig. 4: True target trajectory (solid line) and estimated trajectory (circles) using fixed-interval DRS ($\lambda_y = 0.01, \lambda_x = 0.05$).

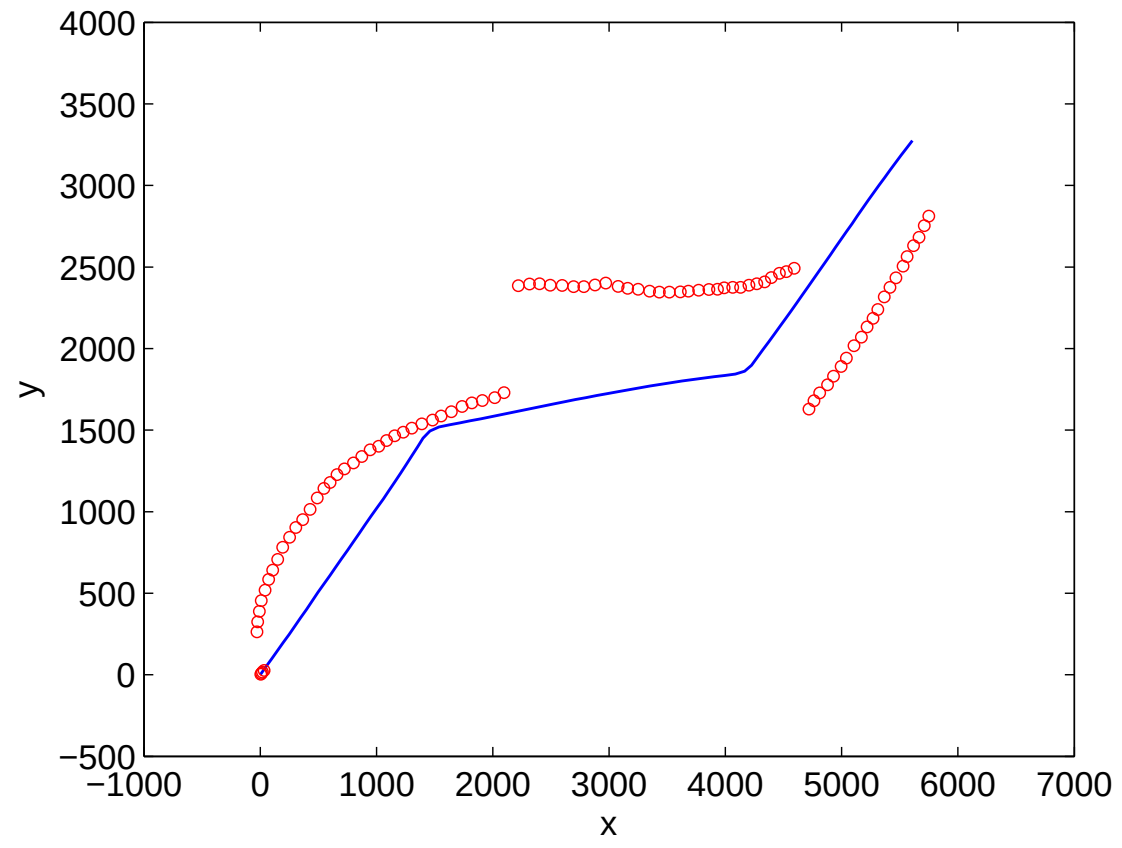

Fig. 5: True target trajectory (solid line) and estimated trajectory (circles) using fixed-lag KS.

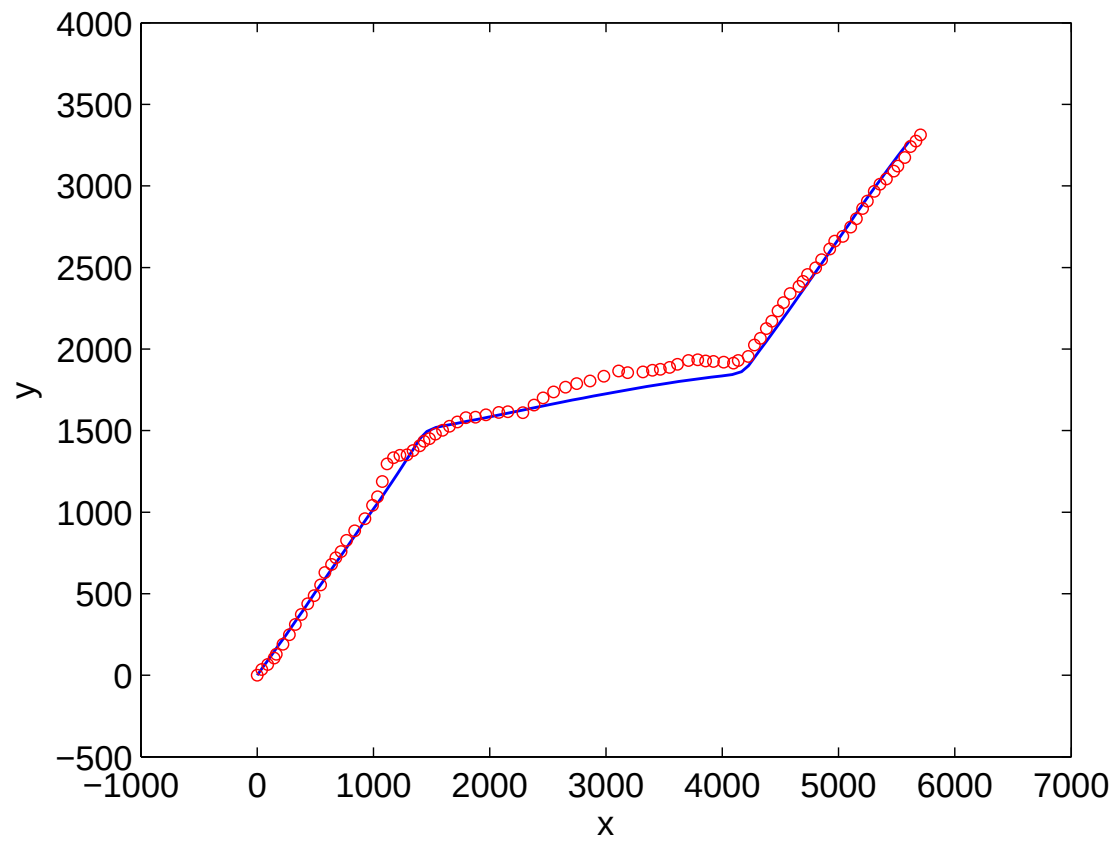

Fig. 6: True target trajectory (solid line) and estimated trajectory (circle) using fixed-lag DRS ($\lambda_y = 0.01, \lambda_x = 0.05$).

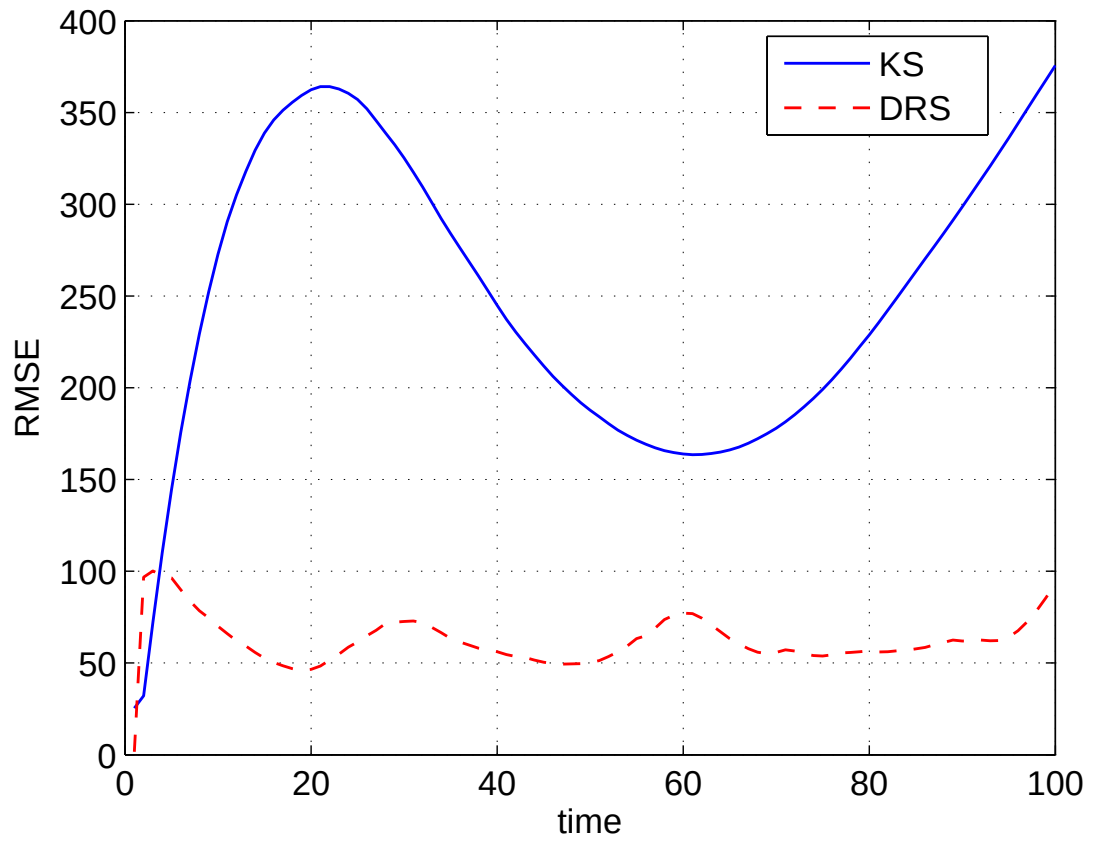

Fig. 7: RMSE analysis of the fixed-interval KS versus DRS ($\lambda_y = 0.01, \lambda_x = 0.05$).

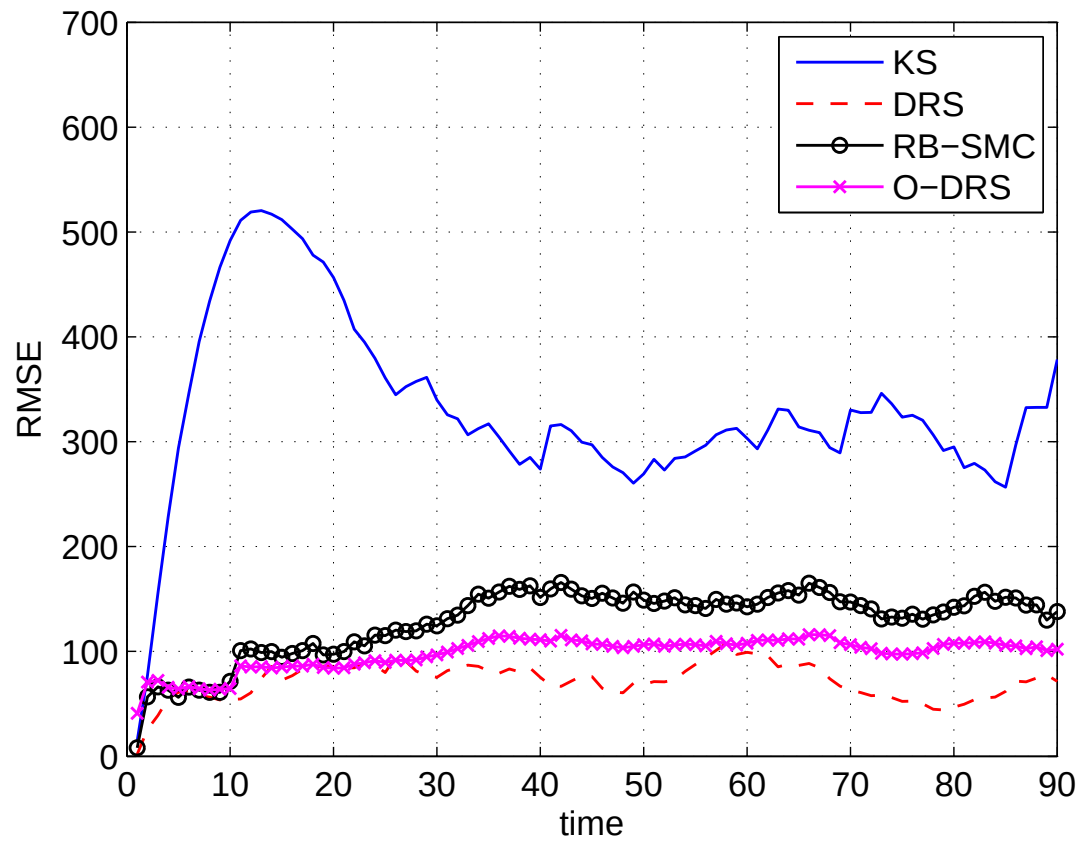

Fig. 8: RMSE analysis of the fixed-lag KS versus DRS ($\lambda_y = 0.01, \lambda_x = 0.05$), online DRS ($\kappa > 0$, $\lambda_y = 0.01, \lambda_x = 0.05$, $J = 50$ AD-MoM iterations), and Rao-Blackwellized SMC smoother (50 particles).

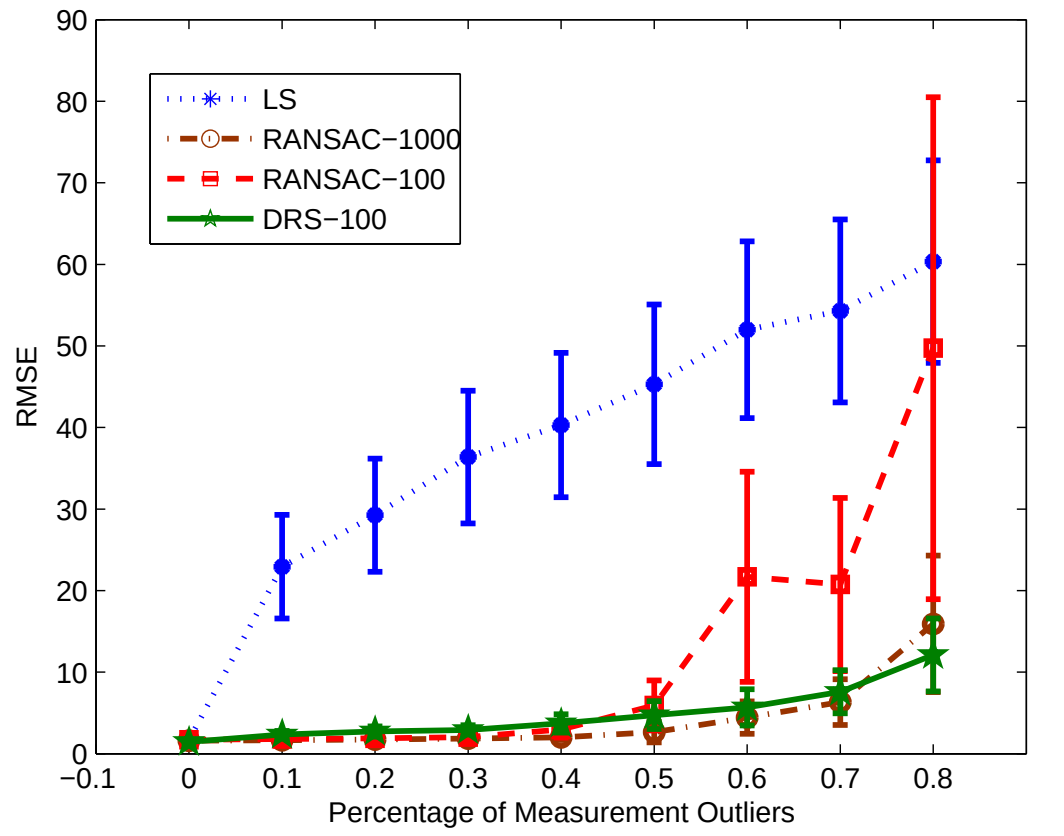

Fig. 9: Mean RMSE $\pm$ std. deviation for estimates formed by RANSAC followed by Huber robustification versus DRS: Measurement outliers only.

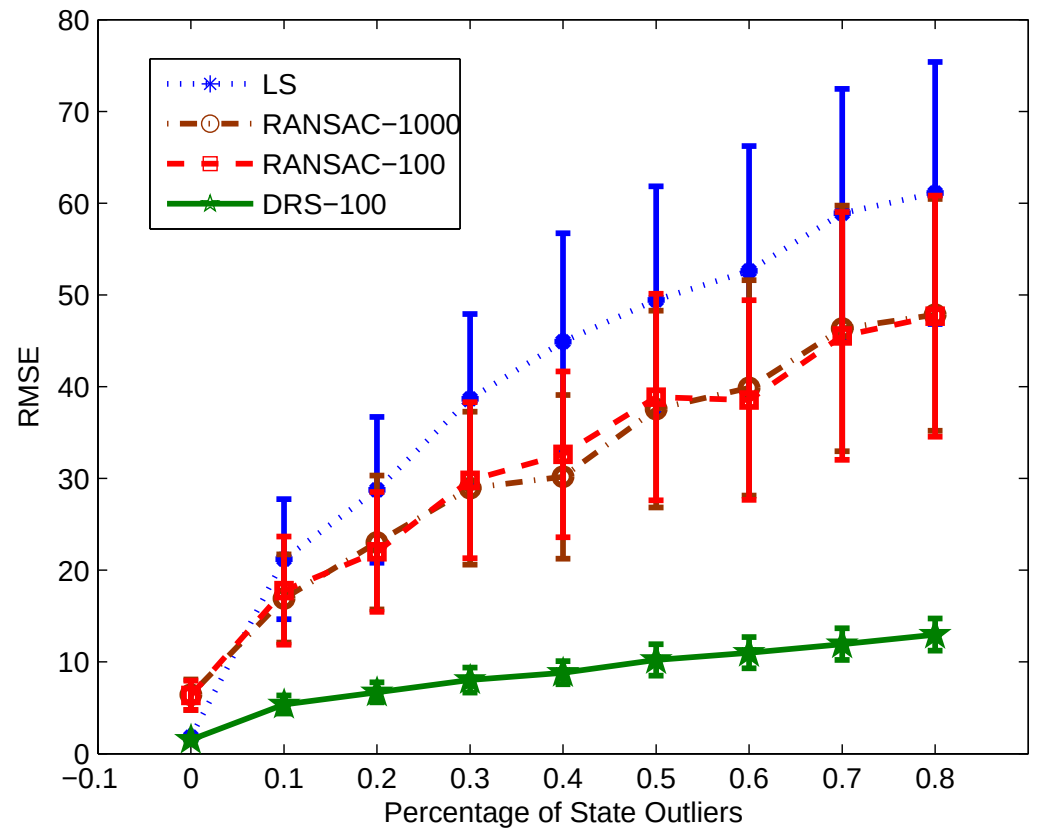

Fig. 10: Mean RMSE $\pm$ std. deviation for estimates formed by RANSAC followed by Huber robustification versus DRS: State outliers only.

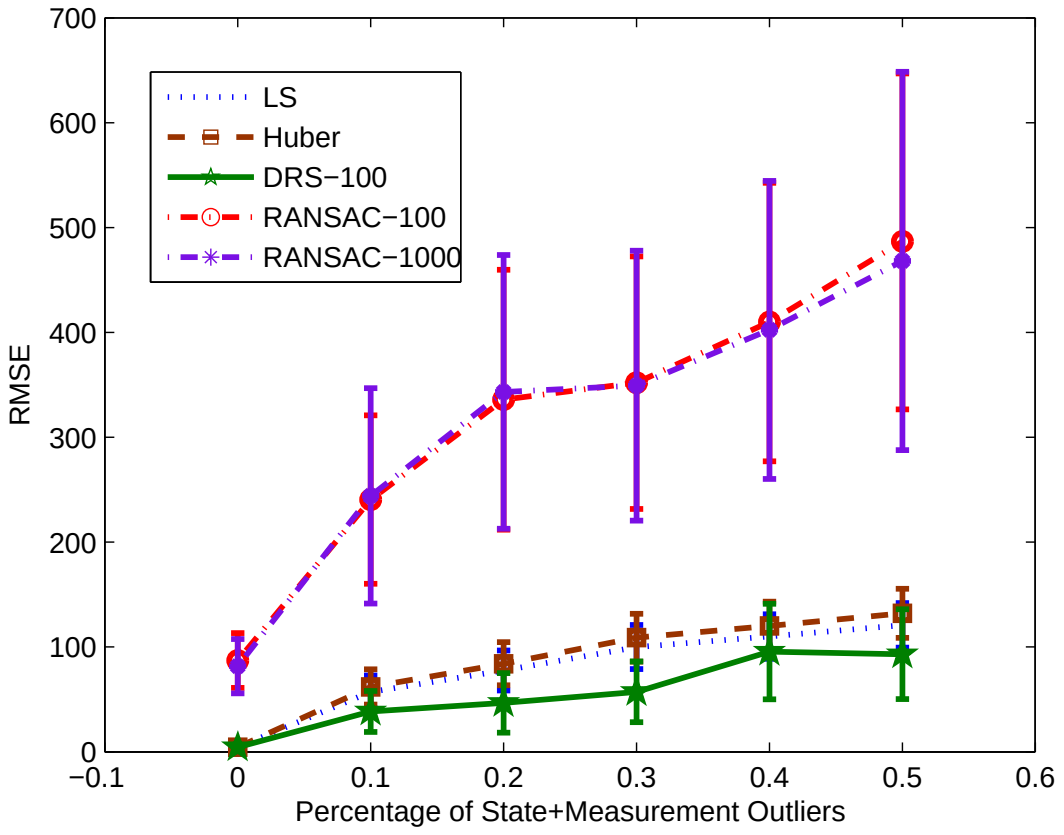

Fig. 11: Outliers present in state and measurements.

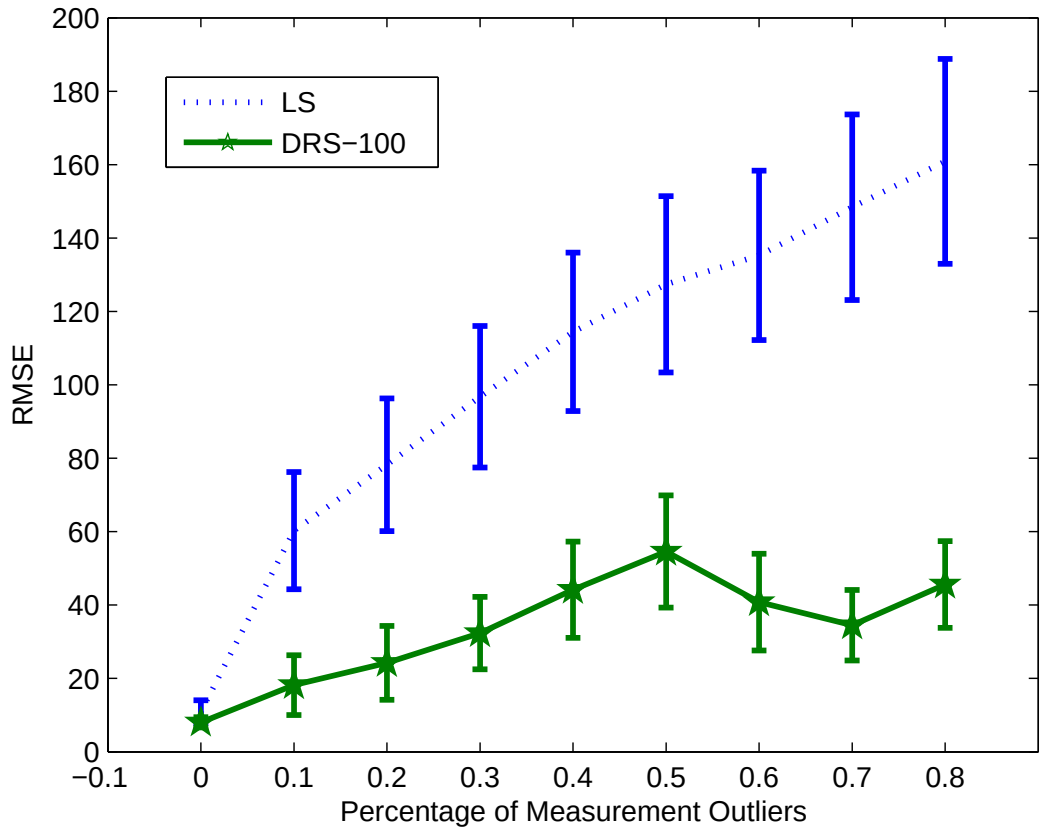

Fig. 12: DRS versus LS with known percentage of outliers.

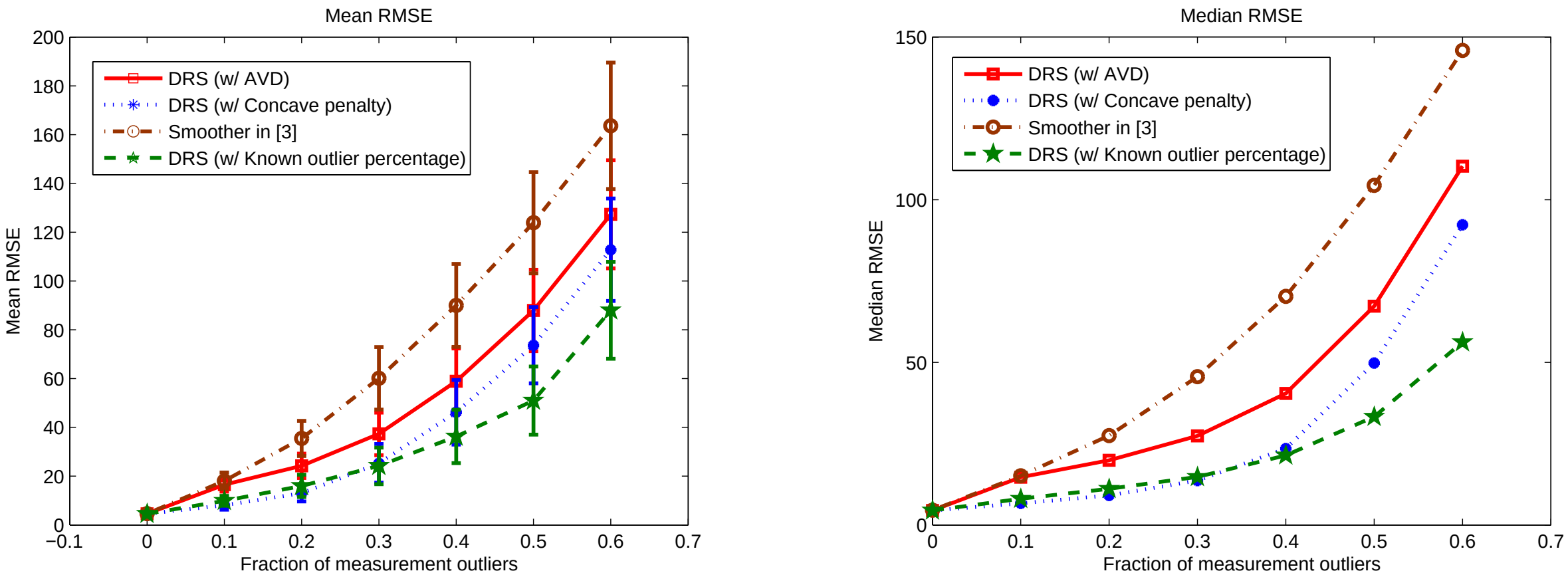

Fig. 13: Comparison among different DRS renditions with the smoother in [3]: (left) Mean RMSE $\pm$ std. deviation; (right) Median RMSE.

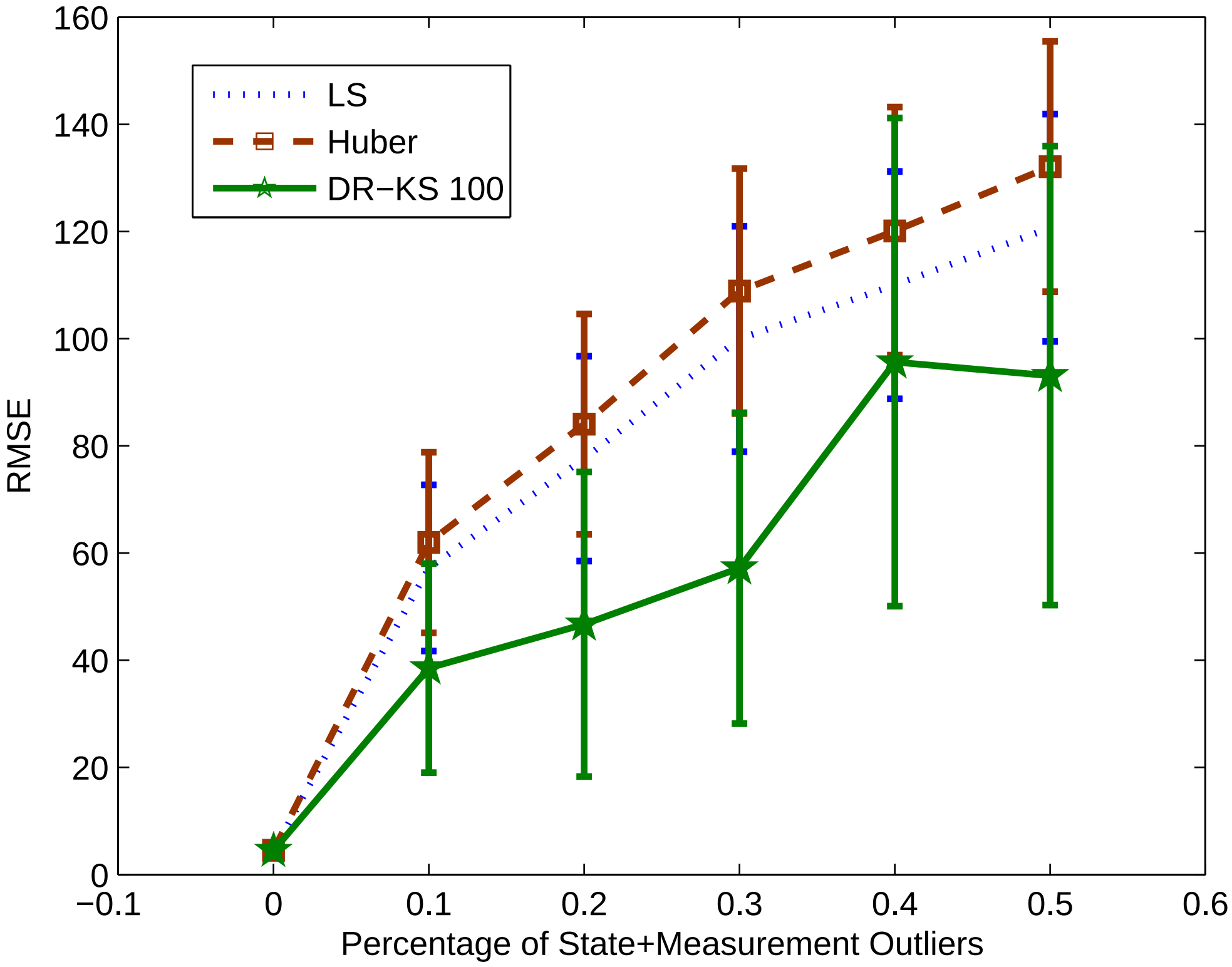